\documentclass[aps,prr,twocolumn,superscriptaddress]{revtex4-1}
\usepackage{amsfonts}
\usepackage{graphicx}
\usepackage{epstopdf}
\usepackage{epsfig}
\usepackage{amsmath}
\usepackage{color}
\usepackage{makecell}
\usepackage{mathtools}

\begin{document}

\title{Scattering of Dirac electrons from a skyrmion: emergence of robust skew scattering}

\author{Cheng-Zhen Wang}
\affiliation{School of Electrical, Computer and Energy Engineering, Arizona State University, Tempe, Arizona 85287, USA}

\author{Hong-Ya Xu}
\affiliation{School of Electrical, Computer and Energy Engineering, Arizona State University, Tempe, Arizona 85287, USA}

\author{Ying-Cheng Lai} \email{Ying-Cheng.Lai@asu.edu}
\affiliation{School of Electrical, Computer and Energy Engineering, Arizona State University, Tempe, Arizona 85287, USA}
\affiliation{Department of Physics, Arizona State University, Tempe, Arizona 85287, USA}

\date{\today}
\begin{abstract}

We study electron scattering from a closed magnetic structure embedded in the top surface of a topological insulator (TI). Outside of the structure there is a uniform layer of ferromagnetic insulator (FMI), leading to a positive effective mass for the Dirac electrons. The mass inside the structure can be engineered to be negative, leading to a skyrmion structure. The geometric shape of the structure can be circular or deformed, leading to integrable or chaotic dynamics, respectively, in the classical limit. For a circular structure, the relativistic quantum scattering characteristics can be calculated analytically. For a deformed structure, we develop an efficient numerical method, the multiple multipole method, to solve the scattering wavefunctions. We find that, for scattering from a skyrmion, anomalous Hall effect as characterized by strong skew scattering can arise, which is robust against structural deformation due to the emergence of resonant modes. In the short (long) wavelength regime, the resonant modes manifest themselves as confined vortices (excited edge states). The origin of the resonant states is the spin phase factor of massive Dirac electrons at the skyrmion boundary. Further, in the short wavelength regime, for a circular skyrmion, a large number of angular momentum channels contribute to the resonant modes. In this regime, in principle, classical dynamics are relevant, but we find that geometric deformations, even those as severe as leading to fully developed chaos, have little effect on the resonant modes. The vortex structure of the resonant states makes it possible to electrically ``charge'' the skyrmion, rendering feasible to manipulate its motion electrically. In the long wavelength regime, only the lowest angular momentum channels contribute to the resonant modes, making the skew scattering sharply directional. These phenomena can be exploited for applications in generating dynamic skyrmions for information storage and in Hall devices.

\end{abstract}

\maketitle

\section{Introduction} \label{sec:intro}

This paper is devoted to studying relativistic quantum scattering of Dirac 
electrons in systems involving magnetism. There are two motivations. Firstly,
quantum scattering of spin-1/2 fermions is fundamental to developing 
two-dimensional (2D) Dirac material based devices. Secondly, magnetic materials
have been efficient carriers of information and the physics of magnetic 
textures has been a topic of significant interest. In general, in quantum 
scattering, the nature of the underlying classical dynamics can play a role. 
For example, consider electronic scattering from a 2D electrical potential 
domain generated by an external gate voltage. In the classical limit of zero 
wavelength, the electrons are point particles and the domain is effectively
a 2D billiard system in which electrons move along straight lines and are
reflected when ``hitting'' the boundary. For a circular domain, the classical 
dynamics are integrable. However, for a deformed domain, e.g., a stadium shaped
domain, the classical dynamics can be ergodic in the phase space. In this 
case, there is sensitive dependence on initial condition because two nearby 
trajectories will diverge from each other exponentially - the hallmark of 
chaos. Since geometric deformations are inevitable in applications, it is
necessary in the study of quantum scattering to take into account the nature
of classical dynamics. Especially, it is useful to consider deformed 
domains to uncover the possible effects of classical chaos on quantum 
scattering.

We employ the setting of a two-dimensional (2D), closed magnetic structure 
embedded in a uniform layer of ferromagnetic insulating (FMI) materials on 
the top of a 3D topological insulator (TI). Outside of the structure, due to 
the FMI layer and the proximity effect, the electrons obey the Dirac equation 
with a positive mass. The mass of the closed structure can be engineered to 
be negative, making it a skyrmion~\cite{nagaosa2013topological,
fert2013skyrmions,everschor2018perspective,ochoa2018colloquium}. The skyrmion
structure can be deformed so that the classical particle motions inside are 
chaotic. The massive Dirac electrons moving on the surface of the TI are 
scattered by the structure. The system thus not only provides a setting for 
exploring new physics associated with scattering of Dirac electrons from a 
magnetic skyrmion for applications (e.g., in spintronics), but also represents 
a paradigm to study the effects of classical chaos on relativistic quantum 
scattering in the presence of magnetism.

To be systematic and general, we consider the cases where the magnetic 
structure on the top of TI can be of either the skyrmion or the non-skyrmion 
type. The structure can simply be a circle, in which case the classical 
dynamics are integrable, or it can be deformed from the circular shape, e.g., 
a stadium, where there is fully developed chaos in the classical limit.
For a circular structure, the various scattering cross sections can be
obtained analytically from the standard partial wave analysis. For a deformed
structure, we adopt an efficient method, the multiple multipole (MMP) method
in optics, to solving the scattering wavefunctions of the two-component Dirac
fermion in the magnetic system. We focus on two regimes: the short wavelength
regime where the size of the magnetic structure is larger than the wavelength
so that the underlying classical dynamics are relevant, and the long wavelength
regime where the structure size is comparable or smaller than the wavelength. 
There are two main results. Firstly, a skyrmion can lead to strong skew 
scattering due to the emergence of resonant modes that manifest themselves as 
confined vortices inside the skyrmion in the short wavelength regime or confined
edge states in the long wavelength regime. The resonant modes are the result
of mass sign change across the skyrmion boundary. For a circular skyrmion,
in the short wavelength regime, a large number of angular momentum channels
contribute to the resonant modes and electron charging arises, providing a way 
to electrically manipulate the skyrmion motion. In the long wavelength regime,
only the lowest angular momentum channels contribute to the resonant states,
leading to strongly directional skew scattering with implications in
developing Hall devices. The second result is that classical chaos generated 
by geometrical deformations has little effect on the scattering from a 
skyrmion. The scattering phenomena uncovered for the circular case are thus 
robust. The immunity of the scattering dynamics to severe deformation of the 
skyrmion structure is advantageous for spintronic device applications. 

This paper is organized as follows. In Sec.~\ref{sec:background}, we provide
the background of our research in terms of magnetic skyrmion, TIs, the 
``marriage'' between skyrmion and TI, and relativistic quantum chaos. 
In Sec.~\ref{sec:model}, we describe the Hamiltonian and outline the 
methods (analytic and numerical). In Sec.~\ref{sec:resonance}, we demonstrate
the emergence of robust resonant states in scattering from skyrmion for
both integrable and chaotic classical dynamics. In Sec.~\ref{sec:PW}, we 
develop a partial-wave decomposition based analysis for resonances associated 
with scattering corresponding to classical integrable dynamics. In 
Sec.~\ref{sec:discussion}, we summarize the main findings and discuss 
experimental feasibility and open issues. Finally, in Appendix, we detail 
the developed MMP method for numerically calculating the 
scattering wave functions associated with deformed domain hosting chaotic 
dynamics in the classical limit.

\section{Background} \label{sec:background}

\paragraph*{Magnetic skyrmion.}
Generally, a skyrmion is a particle-like magnetic excitation with a swirling 
topological 2D spin texture, i.e., the spin at the core and the spin at
the perimeter point are in opposite directions~\cite{nagaosa2013topological,
fert2013skyrmions,everschor2018perspective,ochoa2018colloquium}. The small
size of the skyrmions and the possibility of moving them with electrical
currents of small density ($\sim 10^5$A/m$^2$) make them promising candidates
for spintronic storage or logic devices~\cite{nagaosa2013topological,
fert2013skyrmions}. Skyrmions have been experimentally observed in 
chiral magnets~\cite{muhlbauer2009skyrmion,yu2010real} as a result of the
competition between the Dzyaloshiskii-Moriya (DM) interactions, Heisenberg
exchange, and Zeeman interactions. It has been demonstrated that metallic
skyrmions can be driven by spin transfer torque (STT) from the electric
current~\cite{jonietz2010spin,zang2011dynamics,yu2012skyrmion}. Optical
skyrmion lattices have been achieved in an evanescent electromagnetic
field~\cite{tsesses2018optical}. In addition, the topological spin Hall
effect has been demonstrated in which a pure transverse spin current is
generated from a skyrmion spin texture~\cite{yin2015topological,
denisov2016electron,denisov2017nontrivial,ndiaye2017topological,
denisov2018topological}.

\paragraph*{Topological insulators.}
TIs are quantum materials with surface states residing in the bulk insulating
gap~\cite{hasan2010colloquium,qi2011topological}. The edge states are
topologically protected and are robust against nonmagnetic disorders due to
a strong spin-momentum locking. The electron motions on the surface follow the
2D linear dispersion with a single band-touching Dirac point and are described
by the Dirac equation. In spite of the strong spin-momentum locking, the
surface electronic states are sensitive to magnetic perturbations. That is,
the electrons will be scattered off upon encountering a magnetic structure 
on the surface of the TI. The interaction between the topological surface
states and magnetic materials in a quasi-one dimensional setting has been
studied~\cite{tserkovnyak2012thin,wei2013exchange,katmis2016high} where,
due to the spin-momentum locking, the exchange coupling between the 
magnetization and the surface electronic states can lead to intriguing 
phenomena such as anomalous magnetoresistance and unconventional transport
behaviors~\cite{yokoyama2010anomalous,wu2010electron}. The interaction can
also lead to nonlinear or even chaotic dynamics in the evolution of
magnetization of the FMI~\cite{wang2016nonlinear,wang2018emergence}.
For example, complicated dynamics can emerge in the magnetization switching
due to a Hall-current-induced effective anisotropic
field~\cite{garate2010inverse,yokoyama2010theoretical,yokoyama2011current,
tserkovnyak2012thin} and steady self-oscillations can arise in an FMI/TI
heterostructure~\cite{semenov2014voltage,duan2015nonlinear,ndiaye2017dirac}.
A quite recent computational study has revealed phase locking in the 
magnetization dynamics of two FMIs on the top of a 3D TI~\cite{wang2018phase}.

\paragraph*{Skyrmion and TI.}
Efforts in improving thermal efficiency and better manipulating skyrmions
have led to the ``marriage'' between skyrmion and TI, where skyrmions arise
on the surface of a TI. Electric charging of magnetic vortices on the surface 
of a TI was investigated~\cite{nomura2010electric}, and the confinement state 
in the skyrmion structure on the surface of a TI was discovered, paving the 
way to driving skyrmion motion using an applied electric 
field~\cite{hurst2015charged}. Electron skew scattering induced by the 
skyrmion structure on the TI surface was also studied~\cite{araki2017skyrmion}.
Quite recently, the combination of two skyrmions with opposite winding 
numbers, called skyrmionum in an FMI/TI heterostructure was observed in the 
physical space~\cite{Zhangetal:2016,PMKGSS:2018,GSBMP:2019}. Theoretically, 
fluctuation-induced N{\'e}el and Bloch skyrmions on the surface of a TI have 
been predicted~\cite{nogueira2018fluctuation}.

Previous studies focused on scattering of electrons from radially symmetric
skyrmion structures. Deformed skyrmion structure has been studied in recent
years. For example, it was found that Majorana modes are robust against 
skyrmion deformations~\cite{garnier2019topological}. Quantum engineering of
Majorana fermions in deformed skyrmion structure was also
studied~\cite{rachel2017quantized,mascot2018quantum} and deformed
(elongated) skyrmions were used for stabilization and control of Majorana
bound states in proximity to an s-wave superconductor~\cite{GSK:2018}. Shape
dependent resonant modes have been discovered recently in skyrmions in 
magnetic nanodisks~\cite{liu2018shape}.

\paragraph*{Relativistic quantum chaos.}
The study of the manifestations of classical chaos in relativistic quantum
systems was pioneered by Sir Michael Berry and his collaborator~\cite{BM:1987}
and recently emerged as an interdisciplinary field of
research~\cite{LXHG:2018,HXGL:2018} with applications to Dirac material
systems~\cite{CNN:2011,AKB:2016}. In contrast to the traditional field
of (nonrelativistic) quantum chaos~\cite{Stockmann:book,Haake:book} where
classical chaos often bears strong signatures in the corresponding quantum
systems, such ``fingerprints'' tend to be weakened in the relativistic quantum
counterparts~\cite{XL:2019,HXHL:2019}. For example, in scattering (e.g.,
electronic transport through a quantum dot structure), chaos tends to smooth
out fluctuations in scattering matrix elements, quantum transmission, or
conductance~\cite{BS:1988,JBS:1990,LBOG:1992,MRWHG:1992,Ketzmerick:1996,
YHLP:2012} if the quantum behaviors are governed by the Schr\"{o}dinger
equation. However, in two-dimensional (2D) Dirac materials such as graphene,
strong fluctuations of the quantum scattering characteristics can persist to
certain extent in spite of classical chaos~\cite{YHLG:2011a,YHLG:2012}.
Another example is a 2D deformed ring with a line of magnetic flux through the
center, where Schr\"{o}dinger electrons are localized but Dirac electrons can
keep circulating along the edges of the ring domain, generating a
superpersistent current in spite of fully developed classical chaos in the
domain~\cite{XHLG:2015} - a phenomenon that can be exploited for creating a 
robust relativistic qubit~\cite{XHL:2015}. Quite recently, the weakening of 
the manifestations of chaos in spin-1/2 Dirac fermion systems was
studied~\cite{HXHL:2019} using the approach of out-of-time-ordered
correlator~\cite{LO:1969}. It has also been revealed that, for scattering in 
spin-1 Dirac-Weyl fermion systems, a class of robust resonant modes can emerge 
that defy classical chaos completely~\cite{XL:2019}.

\section{Model and method} \label{sec:model}

\begin{figure}
\centering
\includegraphics[width=\linewidth]{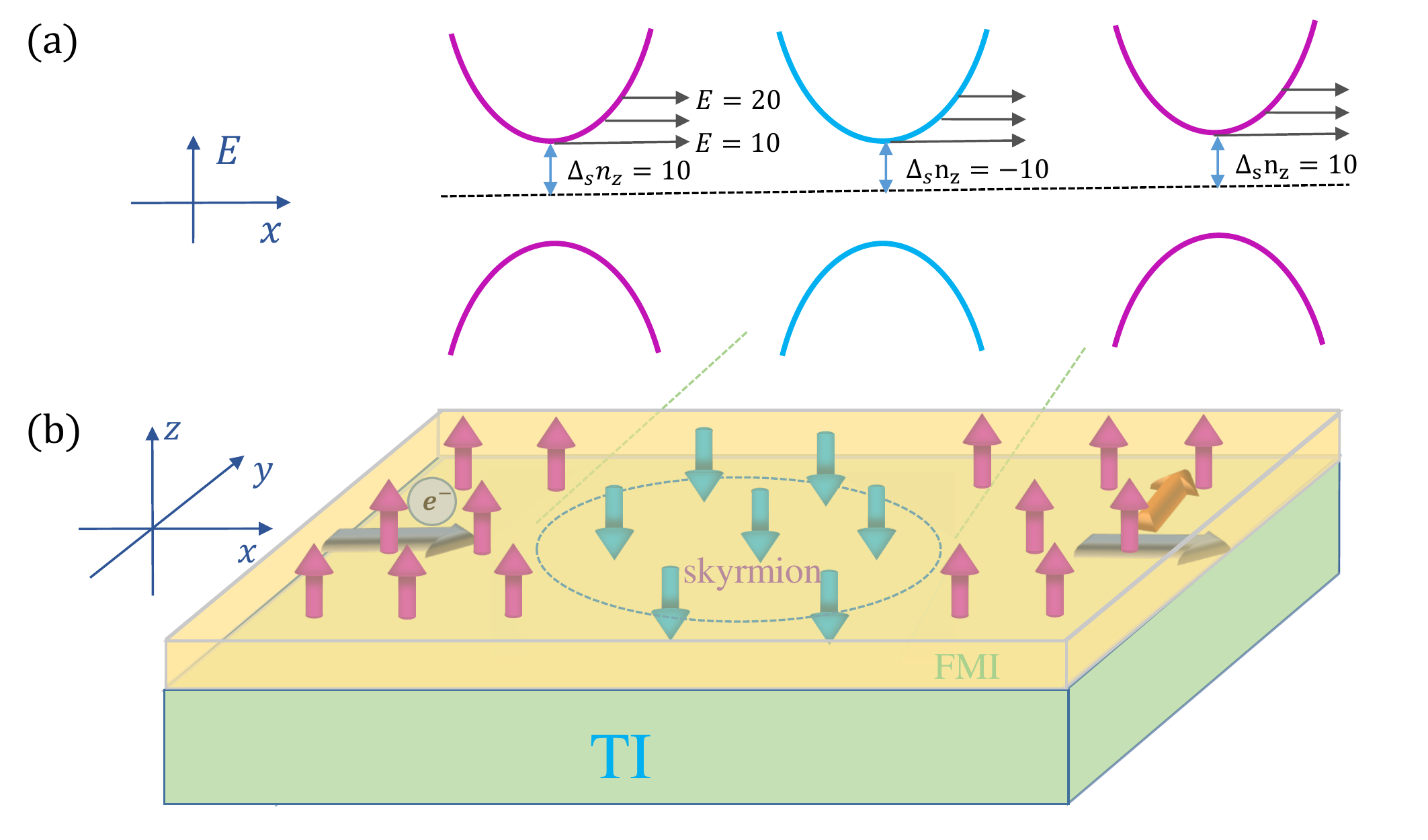}
\caption{ Schematic illustration of electron scattering from a skyrmion
structure in a thin FMI film deposited on the top of a TI. (a) The band
structure of the FMI/TI heterostructure. Outside (inside) of the skyrmion
structure, the mass corresponding to the band gap is positive (negative).
(b) Illustration of electron scattering behavior from the skyrmion structure.
For electronic states outside and inside of the skyrmion, the associated
spin direction is different due to the opposite signs of mass.}
\label{fig:Schematic}
\end{figure}

We place an FMI thin film (e.g., Cu$_2$OSeO$_3$) on the top of a TI with a
single magnetic structure at the center of the thin film, as schematically
illustrated in Fig.~\ref{fig:Schematic}. The motions of the surface electrons
are affected by the structure with the magnetization vector
$\boldsymbol{n}(\boldsymbol{r})$. The Hamiltonian of the system is
\begin{align} \label{Eq:Hamiltonian}
H = v_F (\hat{\boldsymbol{p}}\times \boldsymbol{\sigma})_z
	- \Delta_s \boldsymbol{n}(\boldsymbol{r})\cdot \boldsymbol{\sigma},
\end{align}
where $v_F$ is the Fermi velocity, $\hat{\boldsymbol{p}} = -i\nabla$ is the
momentum operator, $\boldsymbol{\sigma} = (\sigma_x, \sigma_y, \sigma_z)$ are
the Pauli matrices, and $\Delta_s(>0)$ is the spin-splitting energy from the
exchange interaction between the electron and the magnetization. In the polar
coordinates $\boldsymbol{r}=(r,\theta)$, for a circular structure, the
magnetization vector can be parameterized as
\begin{equation}
\boldsymbol{n}(\boldsymbol{r}) = [-\sin{\theta}\sqrt{1 - n_z^2(r)}, \cos{\theta}\sqrt{1 - n_z^2(r)}, n_z(r)].
\end{equation}
For a deformed magnetic structure, there is swirling spin texture with magnetic
moment points up on the edge and down in the center~\cite{schutte2014magnon}.
The out-of-plane component of the magnetic texture $n_z(\boldsymbol{r})$ acts
as a Dirac mass term, which opens a gap in the electronic band structure.
The in-plane component $\boldsymbol{n}_{||}$ can lead to an emergent magnetic
field in the form
\begin{displaymath}
\boldsymbol{B}(\boldsymbol{r}) = \frac{c\Delta \text{div} \boldsymbol{n}_{||}(\boldsymbol{r})}{e\hbar v_F}.
\end{displaymath}	
For a swirling skyrmion structure, the emergent magnetic field $\boldsymbol{B}$
is zero and the in-plane component can be gauged
away~\cite{hurst2015charged,araki2017skyrmion}. In this case, the
``hard-wall'' approximation $n_z(\boldsymbol{r}) = \pm 1$ can be
invoked~\cite{hurst2015charged,araki2017skyrmion}, with the point inside and
outside of the skyrmion structure taking on the value of minus one and one
($n_1 = 1$, $n_2 = -1$), respectively. In experiments, such a structure can
be realized using materials with a strong out-of-plane magnetic anisotropy. In
our study, we assume that the magnetic structure is fixed and unaffected by
the interface electrons. Experimentally, a skyrmion structure can be
stabilized via the Dzyaloshinskii-Moriya (DM) interaction in the
FMI~\cite{nagaosa2013topological,fert2013skyrmions}, where the skyrmion size
depends on  materials parameters such as the relative strength of the
Heisenberg and DM exchange interactions~\cite{nagaosa2013topological,
fert2013skyrmions}. Our model is valid for skyrmion with a vortical magnetic
texture as described. However, for hedgehog skyrmions, the in-plane magnetic
field cannot be gauged away due to the emergent magnetic flux and the
structure is not as stable as vortical skyrmions~\cite{nagaosa2013topological,
fert2013skyrmions}.

The energy-momentum dispersion for electrons in free space with a uniform
magnetic texture (constant mass) is given by
\begin{align} \label{Eq:E-P-Dispersion}
  E_{\pm} = \pm \sqrt{\hbar^2 v_F^2(k_x^2 + k_y^2) + \Delta^2n_z^2},
\end{align}
as shown in Fig.~\ref{fig:Schematic}(a). While the energy dispersion curve
inside of the skyrmion appears similar to that outside of skyrmion, the spin
direction is different for the electronic state due to the opposite signs
of mass. An electron will then go through a scattering process in this
2D system. Because of the breaking of the time reversal symmetry, skew
scattering will arise.

For a circular magnetic structure, the scattering wavefunction and the
related behavior can be solved analytically using the partial-wave
decomposition method (Sec.~\ref{sec:PW}). For a deformed skyrmion,
analytic solutions of the scattering wavefunction are not feasible. We have
developed an MMP based method, which has its origin in
optics~\cite{LB:1987,Imhof:1996,KA:2002,MEHV:2002,TE:2004} and has recently
been extended to scattering of pseudospin-1 particles~\cite{XL:2019}. The
basic idea is to assume two sets of fictitious poles along and in the vicinity
of the entire boundary of the magnetic structure: one outside and another
inside of the boundary. Each pole emits a wave in the form of Hankel function
(spherical wave in the far field). The transmitted wavefunction at each point
inside of the scatterer can be expressed as the superposition of the waves
emitted by the poles outside of the scatterer. Similarly, the refracted
wavefunction at each point outside of the scatterer can be written as the
combination of the waves emitted by the poles inside of the scatterer. The
incident plane wave as well as the reflected and transmitted waves are matched
on the boundary to enable the poles to be determined, and the expansion
coefficients can be obtained by solving the matrix eigenfunctions. (The
details of the MMP method adopted to scattering from a magnetic structure
are given in Appendix) We validate the method by comparing
the MMP solutions with the analytic solution based on partial wave expansion
for a circular skyrmion. Overall, the MMP method is effective and efficient
for solving both the near- and far-field scattering problem for a magnetic
scatterer of {\em arbitrary} shape.

In our calculation, we use the dimensionless quantity obtained via
considerations of the scales of the physical quantities involved. In particular,
the energy scale in the FMI/TI heterostructure is on the order of meV. In free
space with zero mass, the wavevector corresponding to the energy of $1$ meV
is $k\sim 1 meV/(\hbar v_F) = 3.04 \times 10^{-3}/nm$. We take the
dimensionless radius of the magnetic structure (circular shape) to be $R = 1$,
which corresponds to a real structure of size of $100$ nm. We then set the
dimensionless energy corresponding to $1$ meV to be $kR = 0.304$. For
$\Delta = 10$, the corresponding energy gap is $10/0.304 \approx 33$ meV.

\section{Emergence of robust resonant states in scattering from skyrmion}
\label{sec:resonance}

\subsection{Short wavelength regime - resonant vortices and edge modes}

We concentrate on regime where the wavelength of the incoming Dirac electron
is smaller than the size of the magnetic structure so that the classical
dynamics inside the structure are relevant. We consider a circular structure
as well as a deformed structure that leads to chaos in the classical limit to
identify any effect of chaos on the electron scattering behavior.

\begin{figure} [ht!]
\centering
\includegraphics[width=\linewidth]{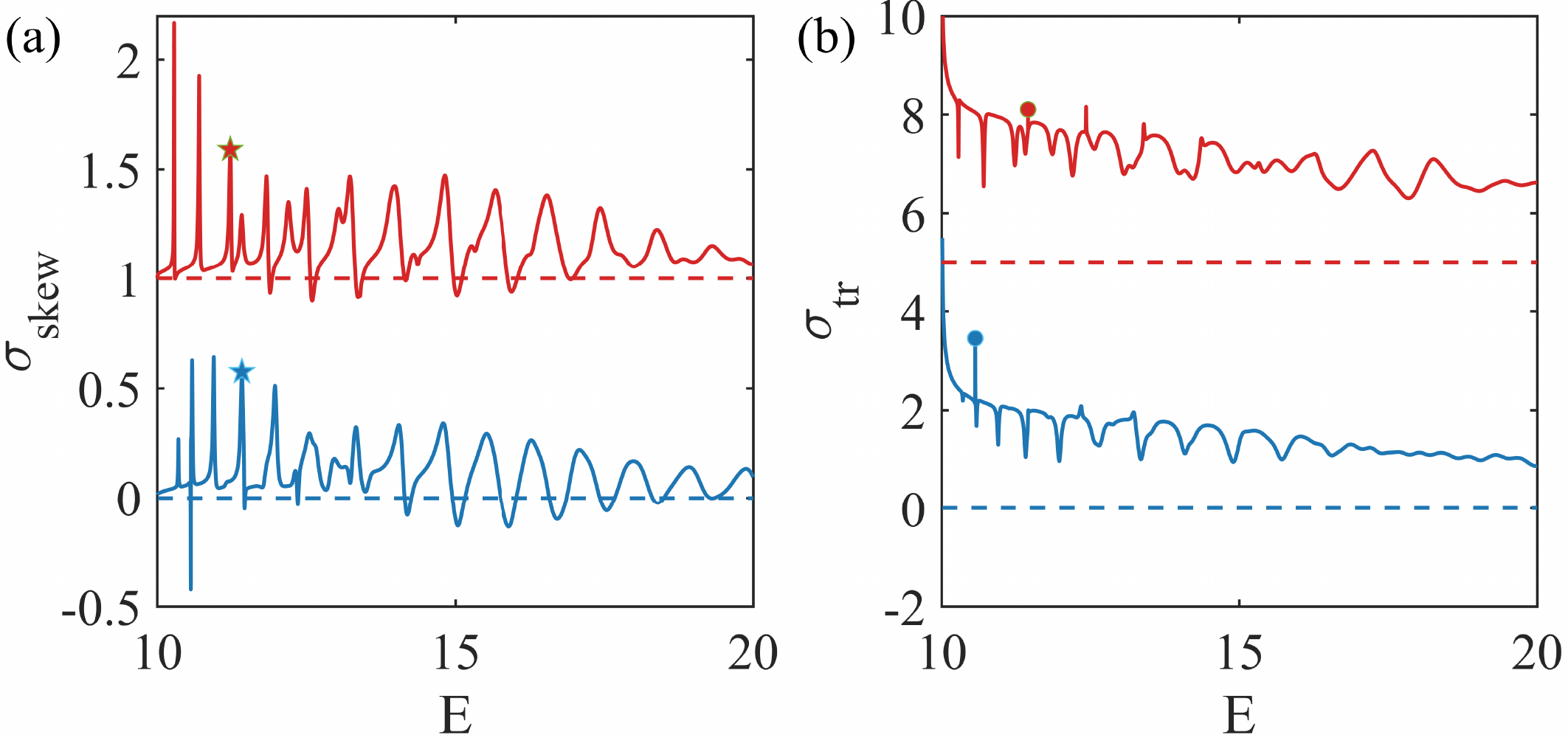}
\caption{ Skew scattering and transport cross sections versus incident
electron energy in the short wavelength regime. (a) Skew scattering cross
section versus the energy. The red and blue curves correspond to a circular
and stadium-shaped skyrmion, respectively. The mass values are $m_1 = 10$ and
$m_2 = -10$.  (b) Backscattering cross section as a function of electron
energy for the two skyrmion shapes as in (a). In each panel, the red curve
has been shifted upwards by an amount specified by the horizontal red-dashed
line for better visualization and comparison with the blue curve.}
\label{fig:CS-LargeR}
\end{figure}

{\em Far-field behavior}. Far away from the scattering center, for unit
incident density the spinor wavefunction can be written as
\begin{align}\label{Eq:wave_outside}
&\Psi_{I} = \Psi_{inc} + \Psi_{ref} \nonumber \\
&\approx C \left(
      \begin{array}{cc}
        1 \\
        i\frac{\hbar v_F k}{E - m_1}
      \end{array}
\right) e^{ikr\cos\theta}  + C
\left(
      \begin{array}{cc}
        e^{-i\theta} \\
        i\frac{\hbar v_F k}{E - m_1}
      \end{array}
\right) \frac{f(\theta)}{\sqrt{r}} e^{ikr}
\end{align}
where $C$ is the normalization factor, $k=\sqrt{k_x^2+k_y^2}$ is the electron
wavevector, $m_1 = \Delta_s n_1$ and $m_2 = \Delta_s n_2$ are the mass terms
outside and inside of the magnetic structure, $f(\theta)$ denotes the 2D
far-field scattering amplitude in the direction defined by angle $\theta$ with
the $x$-axis. For a circular structure, $f(\theta)$ can be obtained
analytically. For a chaotic structure, once the reflection function is
calculated from the MMP method, $f(\theta)$ can be obtained. The differential
cross section is
\begin{align} \label{Eq:DifCS}
  \frac{d\sigma}{d\theta} = |f(\theta)|^2.
\end{align}
The transport and skew cross sections are defined, respectively, as
\begin{align} \label{Eq:BackCS}
  \sigma_{tr} = \int_{0}^{2\pi} d\theta |f(\theta)|^2(1 - \cos\theta)
\end{align}
and
\begin{align} \label{Eq:SkewCS}
  \sigma_{skew} = \int_{0}^{2\pi} d\theta |f(\theta)|^2 \sin\theta.
\end{align}
Figures~\ref{fig:CS-LargeR}(a) and \ref{fig:CS-LargeR}(b) show, respectively,
the skew scattering and transport cross sections as a function of incident
electron energy, for a skyrmion (negative value of $m_2$) of circular shape
(upper panel) and stadium shape (lower panel) of the same area $\pi$ in
dimensionless units. The stadium shape is chosen because of its mirror
symmetry for the incident plane waves so as to avoid an unnecessary
complication: mixing of skew scattering and back-scattering (or reflection).
For both skyrmion shapes, there are sharp resonant peaks in the skew cross
section in the lower energy range close to the gap - an indication of the
emergence of anomalous Hall effect associated with Dirac electron scattering
from the skyrmion. As the incident energy is increased, the peak height is
reduced but its width becomes larger, as a larger energy value corresponds
to less distortion in the energy-momentum dispersion with the mass gap. Note
that there is little difference in the skew scattering cross section curves
for the two skyrmion shapes, indicating that the nature of the classical
dynamics hardly affects the scattering. For the curves of the transport cross
section, as shown in Fig.~\ref{fig:CS-LargeR}(b), its value decreases with
increasing energy. For low energy values, the valleys in the transport cross
section correspond exactly to the skew scattering peaks. Sharp peaks also
exist in the backscattering cross section curve. Similar to the skew cross
section, the nature of the classical dynamics has no appreciable effect. The
results in Fig.~\ref{fig:CS-LargeR} indicate that skyrmion skew scattering is
robust against geometric deformations that are so severe as to change the
classical behavior completely: from integrable dynamics to chaos.

\begin{figure} [ht!]
\centering
\includegraphics[width=\linewidth]{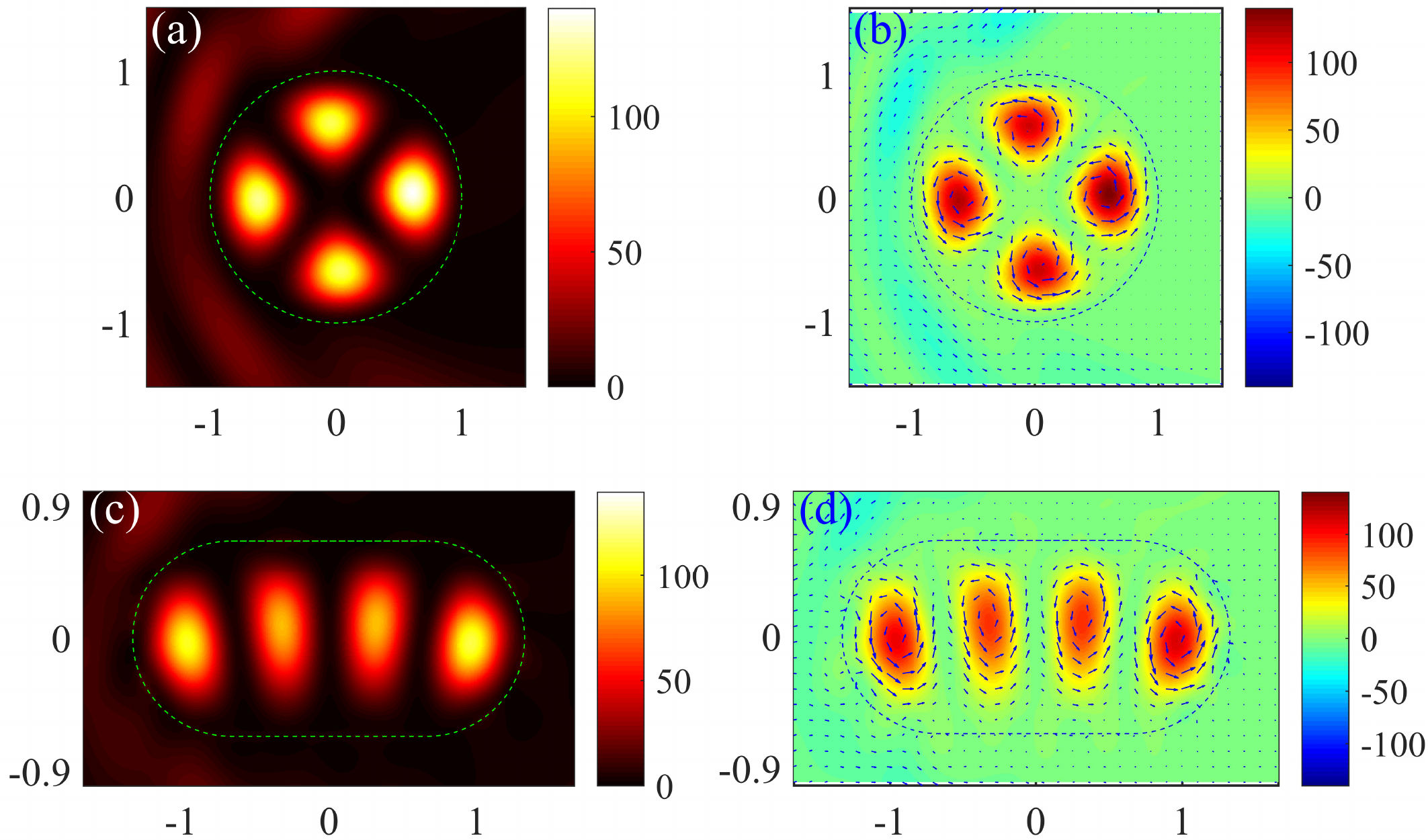}
\caption{Probability and current density distribution for selected vortex
states. (a) The probability distribution for scattering from a circular
skyrmion for $m_1 = 10$, $m_2 = -10$, and $E = 11.225$. (b) In-plane current
(marked as arrows) and spin-$z$ component (color coded) density distribution
in the circular skyrmion region. (c,d) The corresponding probability, current
and spin distribution for scattering from a stadium-shaped skyrmion for
$m_1 = 10$, $m_2 = -10$, and $E = 11.42$.}
\label{fig:State-LargeR}
\end{figure}

{\em Near-field behavior}. To understand the origin of the deformation (chaos)
independent far-field scattering (transport) behavior, we study the near-field
scattering behavior by examining the probability density and the current
density distribution associated with some specific energy state. In particular,
the probability density is given by $P = \Psi^\dagger \Psi$, where
$\Psi=(\psi_1,\psi_2)^T$ is the wavefunction, and the probability current
operator is $\hat{J} = \nabla_{\boldsymbol{p}}H=v_F(\sigma_y, -\sigma_x)$.
The current density can be obtained as
\begin{align}\label{Eq:Current}
\boldsymbol{J}=(J_x,J_y)=v_F[2\real(i\psi_1\psi_2^*),-2\real(\psi_1\psi_2^*)].
\end{align}
The probability density distribution of the spin-$z$ component is given by
\begin{displaymath}
\langle \sigma_z \rangle = |\psi_1|^2 - |\psi_2|^2.
\end{displaymath}
We choose a representative energy value corresponding to a skew scattering
cross section peak: $E = 11.225$ for the circular skyrmion and $E = 11.42$
for the stadium-shaped skyrmion - marked as the red and blue stars in
Fig.~\ref{fig:CS-LargeR}(a), respectively. The probability and the
current density distributions are shown in Fig.~\ref{fig:State-LargeR}.
From both skyrmion structures, there are scattering resonant states, as shown
in Figs.~\ref{fig:State-LargeR}(a) and \ref{fig:State-LargeR}(c). The resonant
patterns correspond to weak backscattering but stronger skew scattering cross
sections, indicating that these are effectively quasi-confined states. Further
insights into the contribution of the resonant states to skew scattering can
be gained by examining the current density distribution (marked as arrows)
and the spin-$z$ component density distribution (color coded) in the 2D
skyrmion structure, as shown in Figs.~\ref{fig:State-LargeR}(b) and
\ref{fig:State-LargeR}(d). We see that the confined resonant states form
vortices with counter-clockwise currents. There is also an out-of-plane spin
component along the positive $z$ direction. The vortices have an apparent
directionality, so they can affect the skew scattering direction and
magnitude. The vortices are formed by interference of waves reflected from
the boundary and are robust against boundary deformation. As a result, the
nature of the classical dynamics, integrable or chaotic, has no significant
effect on scattering.

\begin{figure}
\centering
\includegraphics[width=\linewidth]{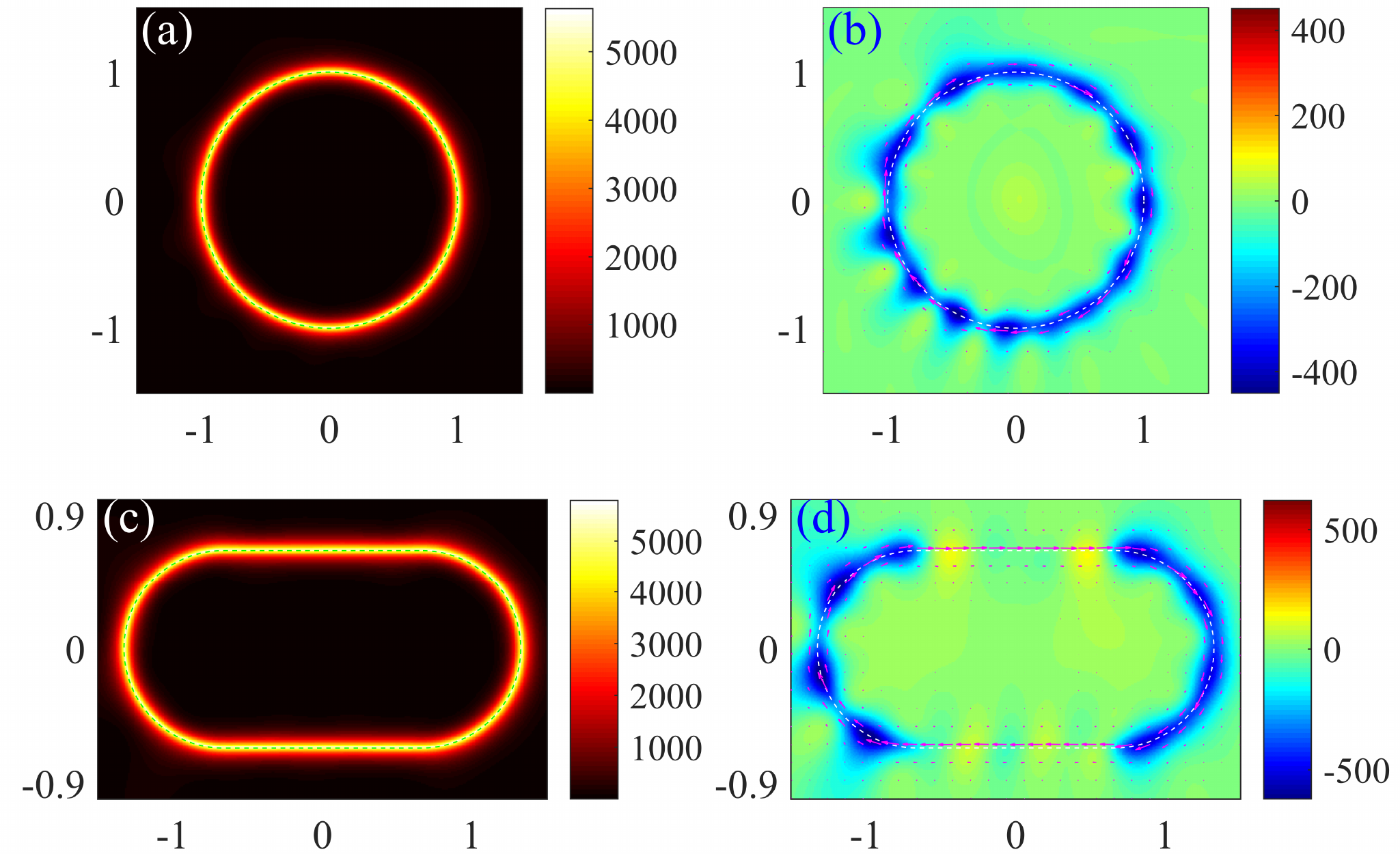}
\caption{ Wavefunction probability and current density distribution associated
with selected edge states. (a) The probability distribution for scattering
from a circular skyrmion for $m_1 = 10$, $m_2 = -10$, and $E = 11.461$. (b)
The corresponding in-plane current (marked as arrows) and spin-$z$ component
(represented by colors) density distribution. (c,d) The probability and spin
distributions associated with scattering from a stadium-shaped skyrmion for
$m_1 = 10$, $m_2 = -10$, and $E = 10.564$.}
\label{fig:Edge-State-LargeR}
\end{figure}

In addition to the confined vortex states inside of the skyrmion structure,
another form of confined states arises along the skyrmion boundary, as shown
in Figs.~\ref{fig:Edge-State-LargeR}(a) and \ref{fig:Edge-State-LargeR}(c),
for scattering from a circular and a stadium-shaped skyrmion, respectively.
There is strong confinement of the scattering wavefunction near the boundary
with clockwise current and spin-$z$ component along the negative $z$ axis
direction, as shown in Figs.~\ref{fig:Edge-State-LargeR}(b) and
\ref{fig:Edge-State-LargeR}(d). The edge states correspond to sharp resonant
peaks in the backscattering cross section marked as the filled circles in
Fig.~\ref{fig:CS-LargeR}(b). For the circular skyrmion, the edge states have
no corresponding sharp peaks in skew scattering. For the stadium-shaped
skyrmion, the edges states correspond to sharp valleys in the skew scattering
cross section.

\subsection{Long wavelength regime - resonant modes near the boundary}

\begin{figure} [ht!]
\centering
\includegraphics[width=\linewidth]{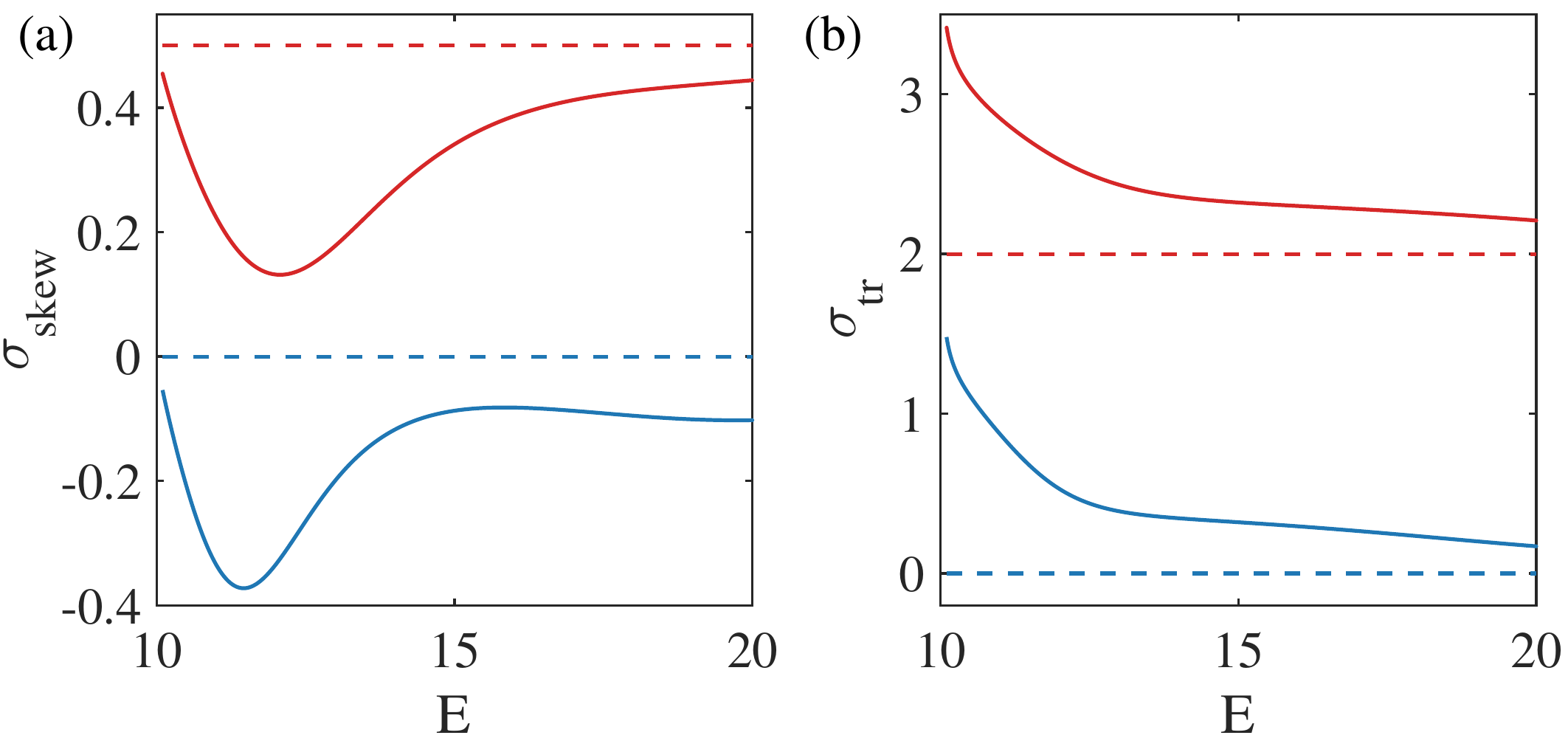}
\caption{ Characteristics of Dirac electron scattering from a magnetic
skyrmion in the long wavelength regime. (a,b) Skew scattering and
backscattering cross sections versus energy, respectively. The red and blue
curves correspond to a circular and stadium-shaped skyrmion, respectively. The
mass values are $m_1 = 10$ and $m_2 = -10$. In each panel, the red curve has
been shifted upward for a proper amount for better visualization and
comparison with the blue curve.}
\label{fig:CS-SmallR}
\end{figure}

{\em Far-field behavior}. We consider the regime where the skyrmion size
is smaller than the electronic wavelength: $R \ll 1/k$. This can be realized
by setting the area of the skyrmion structure to be $0.01\pi$ for both
circular ($R=0.1$) and stadium-shaped skyrmions. In this long wavelength
regime, for a deformed skyrmion structure, the MMP method is still
effective for calculating the far-field cross sections and the near-field
state distribution. Representative results on the skew scattering and
transport cross sections versus the incident energy are shown in
Fig.~\ref{fig:CS-SmallR}. Different from the scattering behaviors in the
short wavelength regime, the oscillations of the skew scattering cross
section with energy are weak. For example, in the energy range $10<E<20$,
only one smooth peak appears. There is hardly any difference in the
scattering characteristics between the two skyrmion structures, which is
understandable as any structural differences are not resolved in the long
wavelength regime. Because of lack of appreciable oscillations, there is
directional skew scattering over a large energy range - a desired feature
in Hall device applications.

\begin{figure} [ht!]
\centering
\includegraphics[width=\linewidth]{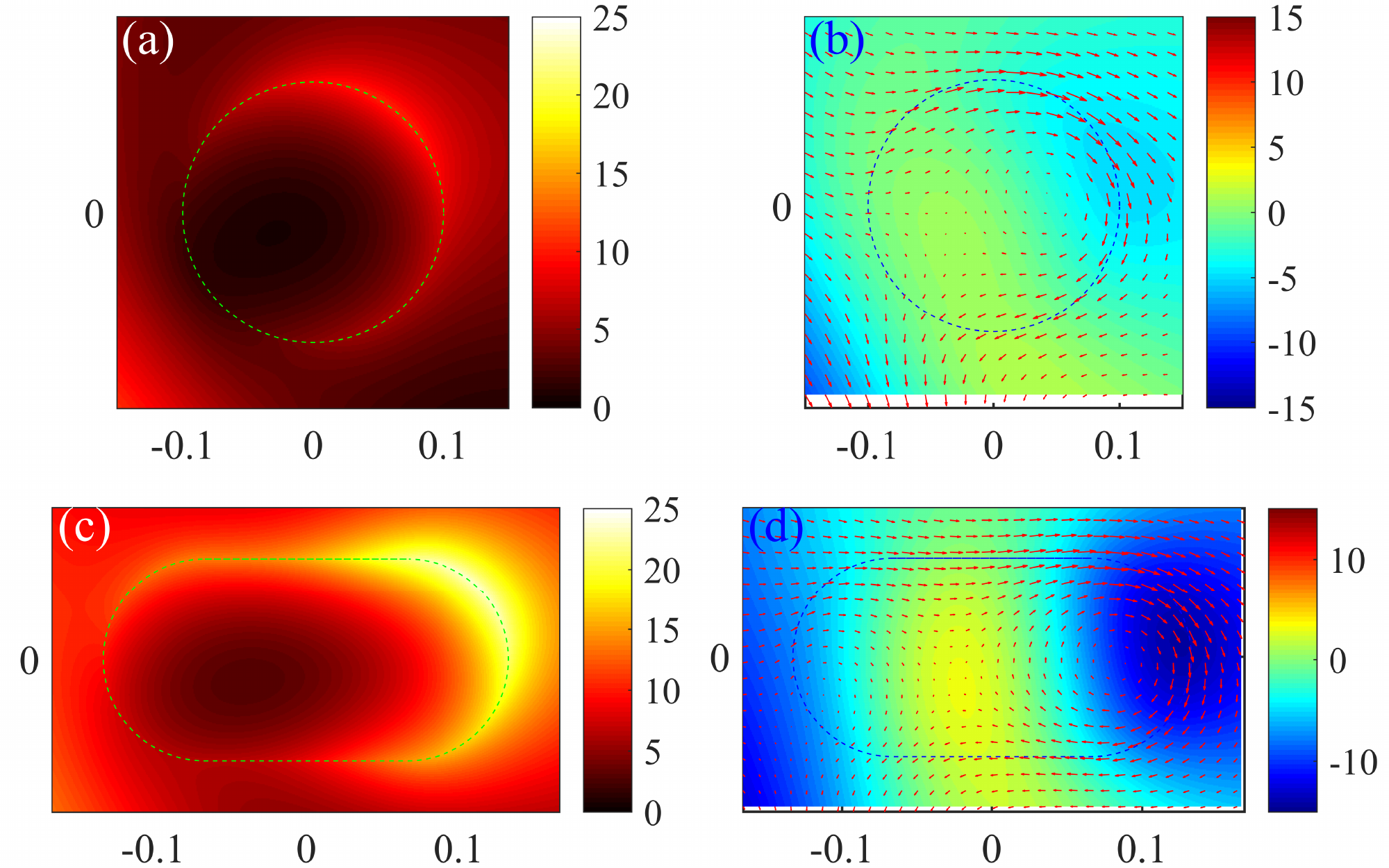}
\caption{ Wavefunction probability and current density distributions for
selected states for scattering in the long wavelength regime. (a,b) The
probability distribution and in-plane current together with the spin-z
component density distributions, respectively, for scattering from a circular
skyrmion for $m_1 = 10$, $m_2 = -10$, and $E = 12.072$. (c,d) The
corresponding results for scattering from a stadium-shaped skyrmion for
$m_1 = 10$, $m_2 = -10$, and $E = 11.46$.}
\label{fig:State-SmallR}
\end{figure}

{\em Near-field behavior}. We examine the state associated with the energy
value that leads to the lowest skew scattering cross section: $E = 12.072$
for the circular and $E = 11.46$ for the stadium-shaped skyrmion, and the
respective probability density distributions are shown in
Figs.~\ref{fig:State-SmallR}(a) and \ref{fig:State-SmallR}(c). The states
are concentrated in the vicinity of the boundary, which are different from
the vortex states observed in the short-wavelength regime. The edge states
thus represent a different type of resonant states with directional current,
as shown in Figs.~\ref{fig:State-SmallR}(b) and \ref{fig:State-SmallR}(d).
It can be seen that the current direction is downward at the edge,
contributing to skew scattering. The spin-$z$ component is along the negative
$z$ direction.

\subsection{Further demonstration of strong skew scattering from a skyrmion structure}

\begin{figure} [ht!]
\centering
\includegraphics[width=\linewidth]{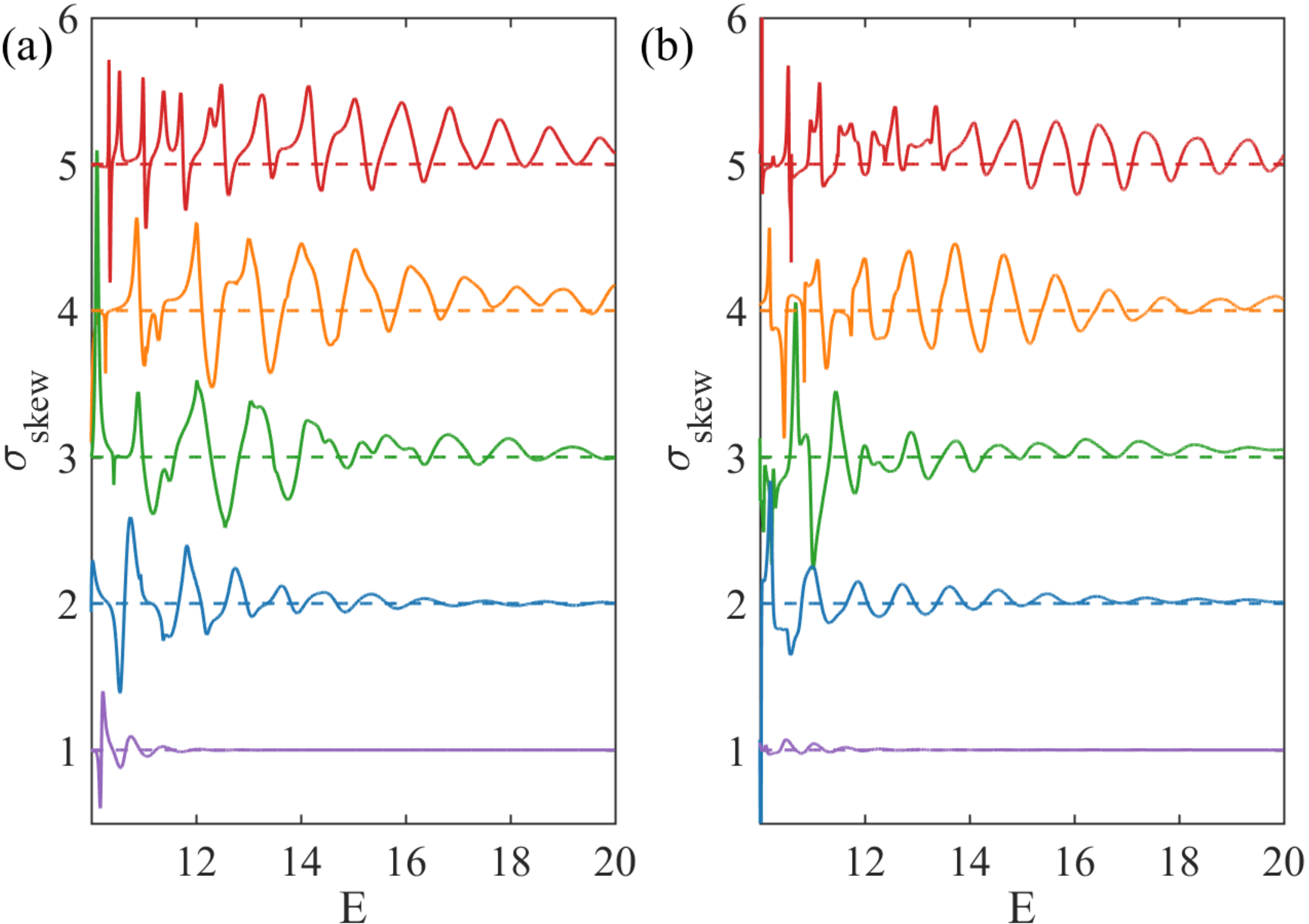}
\caption{ Effects of varying mass on Dirac electron scattering in the
short wavelength regime. The area of the magnetic structure is $\pi$.
(a) Skew scattering cross section versus the electron energy for a circular
structure for mass values $m_2 = -9, -5, 0, 5, 9$, represented by the red,
orange, green, blue and purple solid curves, respectively. In each panel,
the curves have been shifted upward for better visualization and comparison,
where each horizontal dashed line denotes the zero reference point. The mass
outside of the magnetic structure is $m_1 = 10$. (b) The corresponding curves
for a stadium-shape structure with the same mass values as in (a).}
\label{fig:CS-Mass-CircleL}
\end{figure}

To further demonstrate the shape-independent skew scattering behavior of
Dirac electrons from a magnetic structure, we study the effects of changing
the mass of the skyrmion texture. To be concrete, we set $m_1 > 0$ and choose
a set of positive and negative $m_2$ values. In this setting, there is a
skyrmion for $m_2 < 0$ but the magnetic structure is non-skyrmion for
$m_2 > 0$.

\begin{figure}
\centering
\includegraphics[width=\linewidth]{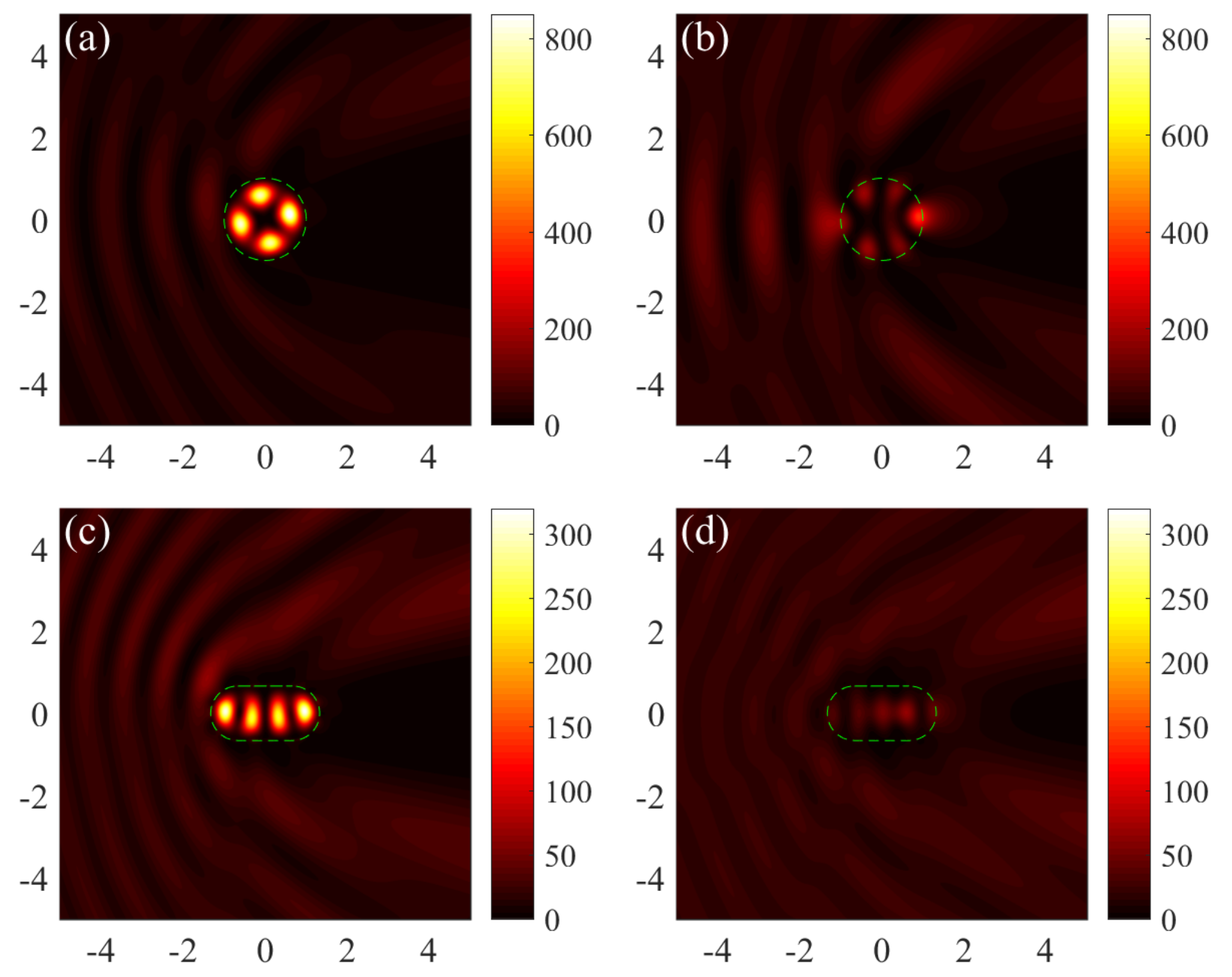}
\caption{ Probability density distribution for selected states in the circular
and stadium-shaped structure for different masses in the short wavelength
regime. (a) Circular skyrmion structure ($m_2 = -9$) for $E = 10.349$,
(b) circular non-skyrmion structure ($m_2 = 9$) for $E = 10.234$,
(c) stadium-shaped skyrmion ($m_2 = -9$) for $E = 10.552$, and
(d) stadium-shaped structure ($m_2 = 9$) for $E = 10.514$.}
\label{fig:CS-Mass-Circle-StateL}
\end{figure}

We first examine the short-wavelength regime to probe into the origin of the
emerged confined vortex states. Figures~\ref{fig:CS-Mass-CircleL}(a) and
\ref{fig:CS-Mass-CircleL}(b) show the skew scattering cross sections for the
circular and stadium-shaped magnetic structure, respectively, for $m_1 = 10$
and $m_2 = -9, -5, 0, 5, 9$. It can be seen that, among the five cases, the
resonant oscillations of the cross section with energy last longer for
$m_2 = -9$. On the contrary, for $m_2 = 9$ (non-skyrmion), the oscillations
diminish rapidly as the energy is increased. These behaviors hold regardless
of whether the underlying classical dynamics are integrable or chaotic.
Overall, a large difference between the masses inside and outside of the
magnetic structure can lead to stronger and long-lasting resonant modes and,
consequently, to more pronounced skew scattering.
Figures~\ref{fig:CS-Mass-Circle-StateL}(a) and
\ref{fig:CS-Mass-Circle-StateL}(b) show the probability density distribution
for $m_2 = 9$ and $m_2 = -9$, respectively, for the circular magnetic
structure. The corresponding results for the stadium-shaped structure are
shown in Figs.~\ref{fig:CS-Mass-Circle-StateL}(c) and
\ref{fig:CS-Mass-Circle-StateL}(d). For both structures, there are resonant
modes for $m_2 = -9$ (when the magnetic structure is of the skyrmion type)
but not for the case of $m_2=9$.

\begin{figure} [ht!]
\centering
\includegraphics[width=\linewidth]{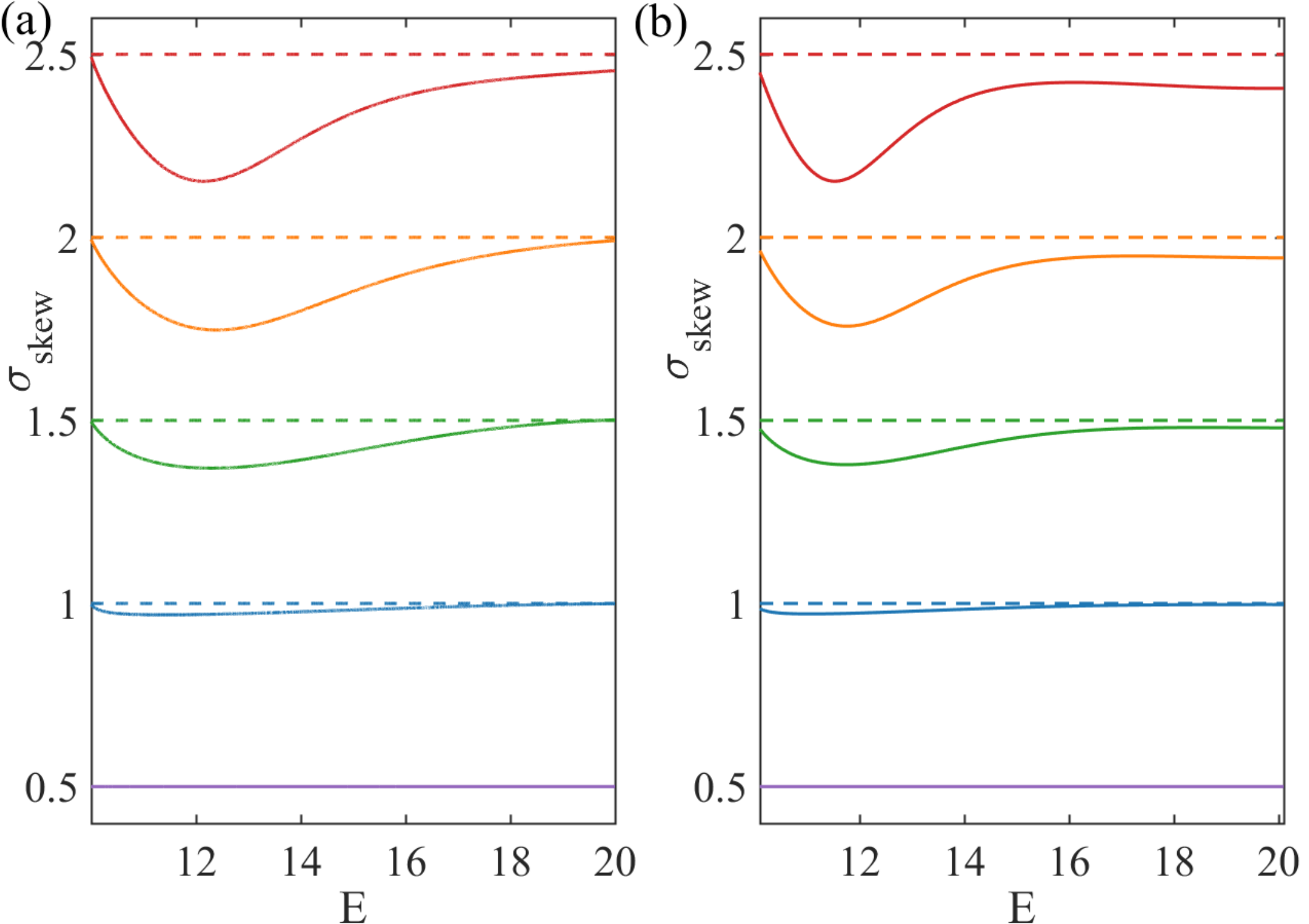}
\caption{ Skew scattering for different mass values of the magnetic
structure in the long wavelength regime. The area of the structure is
$\pi/100$ and the mass outside of the structure is $m_1 = 10$. (a) For a
circular structure, skew scattering cross section for $m_2=-9,-5,0,5,9$,
represented by the red, orange, green, blue and purple solid curves,
respectively. In each panel, the curves have been shifted upward for better
visualization and comparison, with the horizontal dashed lines denoting the
zero reference point. (b) The corresponding results for a stadium-shaped
magnetic structure.}
\label{fig:CS-Mass-CircleS}
\end{figure}

\begin{figure} [ht!]
\centering
\includegraphics[width=\linewidth]{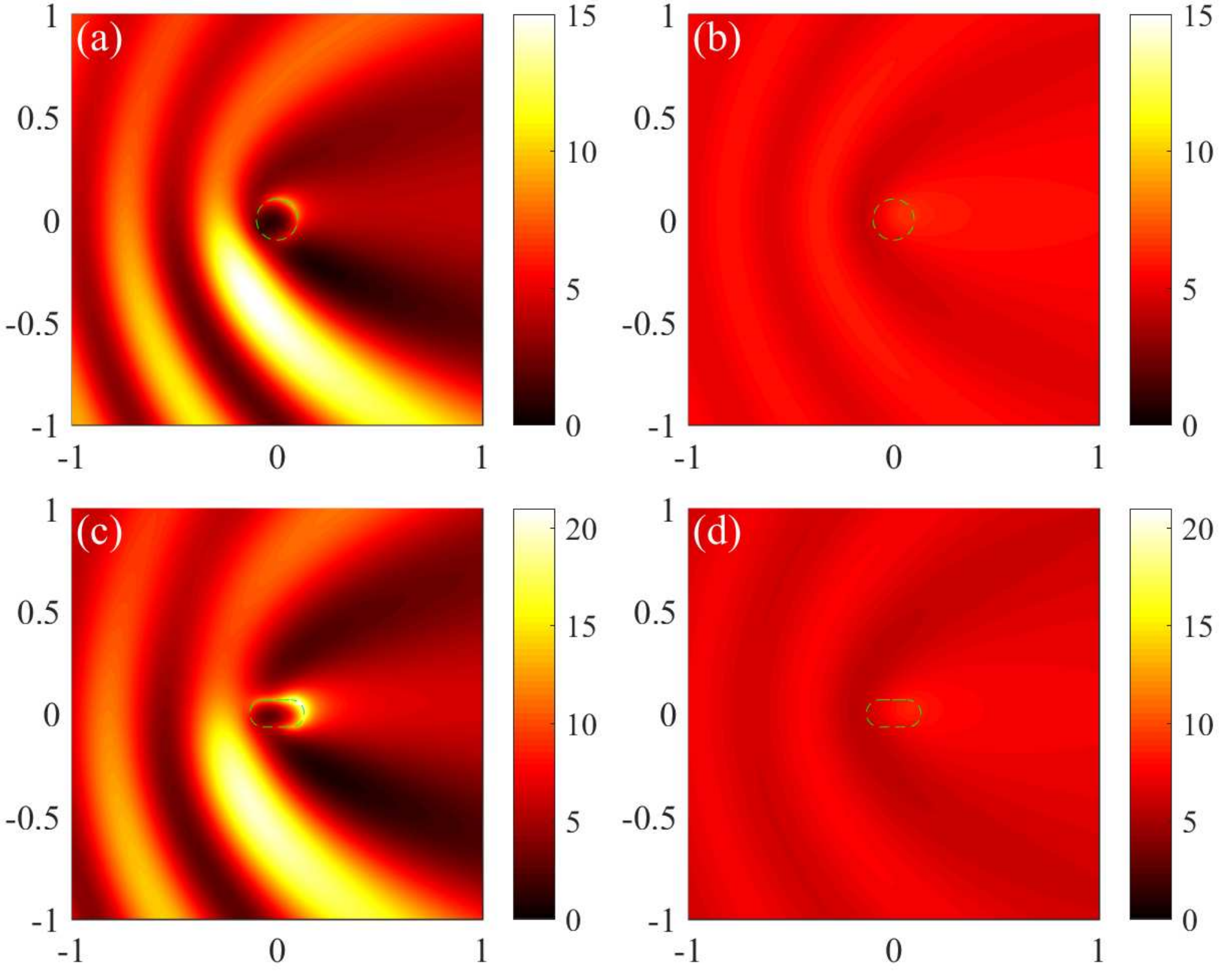}
\caption{ Probability density distribution for the states corresponding to
the minimum of the skew scattering cross section in circular and
stadium-shaped magnetic structures in the long wavelength regime: (a) a
circular skyrmion structure for $m_2 = -9$ and $E = 12.152$, (b) a circular
non-skyrmion structure for $m_2 = 9$ and $E = 12.317$, (c) a stadium-shaped
skyrmion structure for $m_2 = -9$ and $E = 11.53$, and (d) a stadium-shaped
non-skyrmion structure for $m_2 = 9$ and $E = 11.72$.}
\label{fig:CS-Mass-Circle-StateS}
\end{figure}

In the long wavelength regime, regardless of the shape of the magnetic
structure (circular or stadium-shaped), the skew scattering cross section
decreases as the relative mass difference is reduced, as shown in
Fig.~\ref{fig:CS-Mass-CircleS} for $m_2 = -9, -5, 0, 5, 9$.
Figure~\ref{fig:CS-Mass-Circle-StateS} shows representative resonant states
for the circular and stadium-shaped structure for $m_2 = \pm 9$. Again,
when the magnetic structure is of the skyrmion type, skew scattering
is strong, making the scattering electrons directional. However,
when the structure is not of the skyrmion type, skew scattering is weak.

\section{Partial-wave decomposition based analysis} \label{sec:PW}

Numerically, we have observed strong skew scattering of Dirac electrons from
a skyrmion structure, which is robust against geometric deformation. We now
provide an analytic understanding of skew scattering based on the method
of partial wave decomposition. Consider a circular skyrmion. Key to pronounced
skew scattering is the resonant modes emerged from the scattering process.
In the short wavelength regime, a large number of angular momentum components
are involved in the scattering, leading to a large number of resonant modes
as the result of various combinations of the angular momentum components,
which are manifested as peaks in the curve of the cross section with the
energy. In the long wavelength regime, typically only a single resonant mode
is dominant, implying the involvement of only the lowest several angular
momentum components. The asymmetric contribution from different angular
momentum channels leads to the observed pronounced skew scattering. Because
the circular and stadium-shaped skyrmion structures generate similar
scattering behavior, the analytic results from the circular skyrmion case
also provides an understanding of the emergence of strong skew scattering
in the stadium-shaped skyrmion.

For a circular skyrmion, the rotational symmetry stipulates conservation of
the total angular momentum $\hat{J}_z$: $[\hat{J}_z, H] = 0$, and the partial
wave component with total angular momentum $j$ $(= \pm 1/2, \pm 3/2, ...)$
in the polar coordinates $(r, \theta)$ can be written as
\begin{equation} \label{Eq:partial-formula}
\psi_j(\boldsymbol{r}) =
\left(
             \begin{array}{lr}
             u_j(r)e^{i(j-1/2)\theta} \\
             v_j(r)e^{i(j+1/2)\theta}
             \end{array}
\right).
\end{equation}
The Hamiltonian in the polar coordinates is
\begin{equation} \label{Eq:Polar}
H=\hbar v_F
\left(
\begin{array}{cc}
-\frac{\Delta_s n}{\hbar v_F} & -e^{-i\theta}\frac{\partial}{\partial r} + e^{-i\theta}\frac{i\partial}{r\partial \theta}  \\
e^{i\theta}\frac{\partial}{\partial r} + e^{i\theta}\frac{i\partial}{r\partial \theta} & \frac{\Delta_s n}{\hbar v_F}
\end{array}
\right).
\end{equation}
Substituting the partial wave form in Eq.~(\ref{Eq:partial-formula}) into
the Hamiltonian Eq.~(\ref{Eq:Polar}) leads to an eigenvalue problem and
consequently to the explicit expression for the partial waves.

The transmitted wave inside of the skyrmion structure ($r<R$) can be
expanded in terms of the partial waves as
\begin{equation}
\psi^T(r, \theta) = C \sum_{l=-\infty}^{\infty} i^{l-1} B_l
\left(
      \begin{array}{cc}
        J_{l-1}(k'r) e^{i(l-1)\theta} \\
        -\frac{\hbar v_F k'}{E - \Delta_s n'} J_l(k' r) e^{il\theta}
      \end{array}
\right),
\end{equation}
and the reflected wave outside of skyrmion ($r>R$) can be written as
\begin{equation}
\psi^R(r, \theta) = C \sum_{l=-\infty}^{\infty} i^{l-1} A_l
\left(
      \begin{array}{cc}
        H_{l-1}(k r) e^{i(l-1)\theta} \\
        -\frac{\hbar v_F k}{E - \Delta_s n} J_l(k r) e^{il\theta}
      \end{array}
\right),
\end{equation}
where $C$ is a normalization factor.
We denote $n$ ($n'$) as the magnetic moment and $k$ ($k'$) as the wavevector
outside (inside) of the skyrmion structure. For the incident electron in the
free region outside of the skyrmion structure, the wavefunction is
\begin{equation}
\psi^{I} = C
\left(
      \begin{array}{cc}
        1 \\
        i\frac{\hbar v_F k}{E - \Delta_s n}
      \end{array}
\right)e^{ikr \cos\theta}.
\end{equation}
Using the Jacobi-Anger identity:
\begin{align} \label{Eq:Jacobi-Anger}
e^{iz\cos\theta} \equiv \sum_{l=-\infty}^{\infty} i^l J_l(z) e^{il\theta},
\end{align}
we can expand the plane wave in the form
\begin{equation}
\psi^{I} = C \sum_{l} i^{l-1}
\left(
      \begin{array}{cc}
        J_{l-1}e^{i(l-1)\theta} \\
        -\frac{\hbar v_F k}{E - \Delta_s n}J_l(kr) e^{il\theta}
      \end{array}
\right).
\end{equation}
Matching the waves at the skyrmion boundary (r = R):
\begin{align} \label{Eq:wave-match}
\psi^I(R) + \psi^R(R) = \psi^T(R),
\end{align}
we get, after some algebraic manipulation,
\begin{align} \label{Eq:Al1}
A_l = \frac{J_{l-1}(kR) J_l(k'R) - \frac{\tau}{\tau'}J_l(kR)J_{l-1}(k'R)}{\frac{\tau}{\tau'}H_l(kR)J_{l-1}(k'R) - H_{l-1}(kR)J_l(k'R)},
\end{align}
and
\begin{align}\label{Eq:Bl1}
B_l = \frac{J_{l-1}(kR) H_l(kR) - J_l(kR)H_{l-1}(kR)}{H_l(kR)J_{l-1}(k'R) - \frac{\tau'}{\tau}H_{l-1}(kR)J_l(k'R)},
\end{align}
where
\begin{eqnarray}
\nonumber
\tau & = & -\frac{\hbar v_F k}{E - \Delta_s n}, \ \mbox{and} \\ \nonumber
\tau' & = & -\frac{\hbar v_F k'}{E - \Delta_s n'}.
\end{eqnarray}
\begin{figure} [ht!]
\centering
\includegraphics[width=\linewidth]{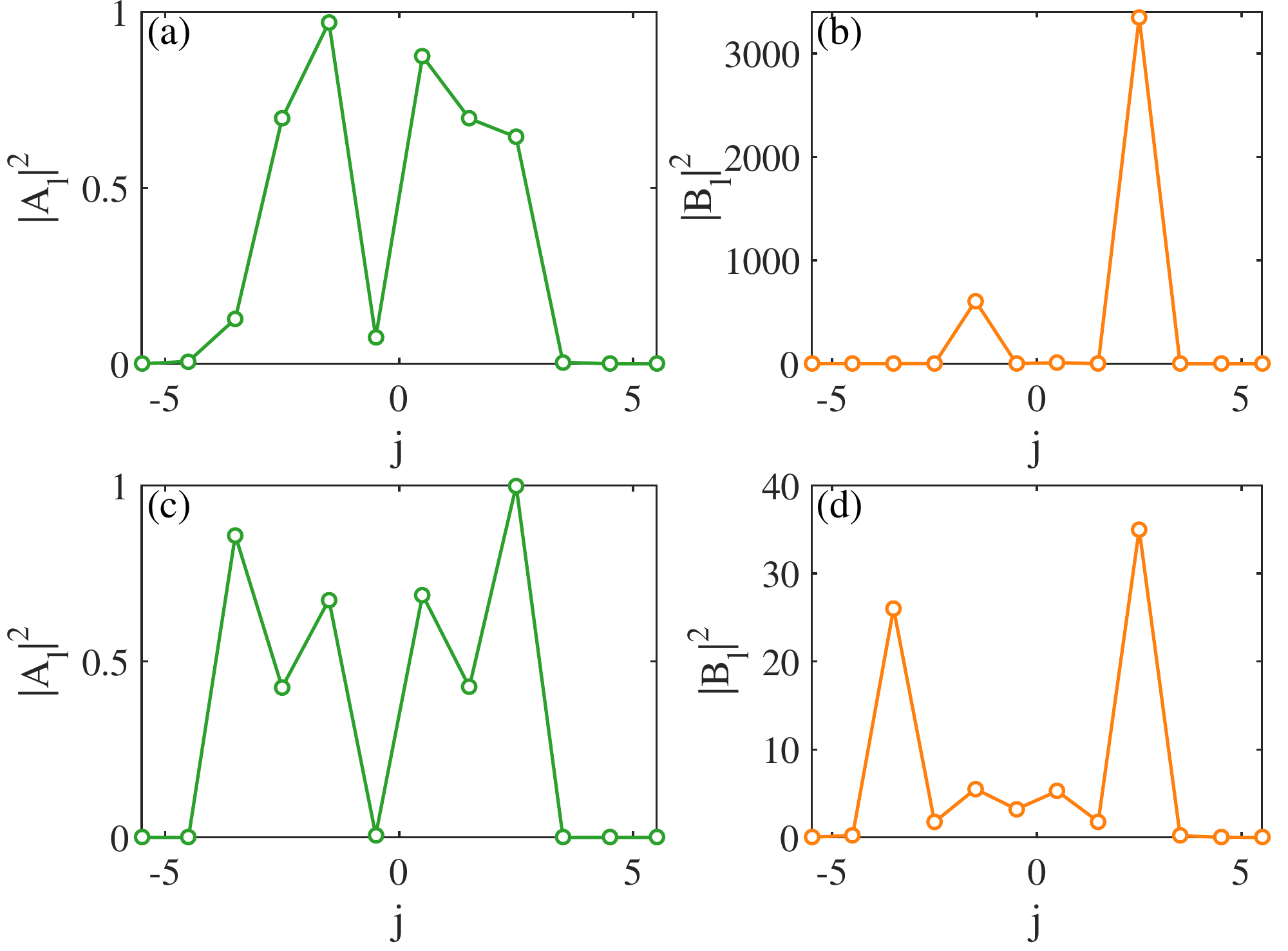}
\caption{ Partial wave decomposition coefficients as a function of total
angular momentum for a circular magnetic structure in the short wavelength
regime. Among the quantities plotted, $A_l$'s are the coefficients for
the reflected waves outside of the structure and $B_l$'s are the transmitted
wave coefficients. (a,b) For a skyrmion structure ($m_1 = 10$ and $m_2 = -9$),
$|A_l|^2$ and $|B_l|^2$ as a function of $j$, respectively, where the
corresponding state is shown in Fig.~\ref{fig:CS-Mass-Circle-StateL}(a).
(c,d) For a non-skyrmion structure ($m_1 = 10$ and $m_2 = 9$),  $|A_l|^2$
and $|B_l|^2$ versus $j$, respectively, where the corresponding state is
shown in Fig.~\ref{fig:CS-Mass-Circle-StateL}(b).}
\label{fig:AB-LargeR}
\end{figure}

Using the explicit formulas for $A_l$ and $B_l$ as given in Eq.~(\ref{Eq:Al1})
and (\ref{Eq:Bl1}), respectively, we obtain the decomposition coefficients
versus the total angular momentum for $R = 1$. Figures~\ref{fig:AB-LargeR}(a)
and \ref{fig:AB-LargeR}(b) show, for the case of scattering from a skyrmion
structure ($m_1 = 10$ and $m_2 = -9$), the expansion coefficients versus
the total angular momentum $j$. Figures~\ref{fig:AB-LargeR}(c)
and \ref{fig:AB-LargeR}(d) show the corresponding results for a non-skyrmion
case ($m_1 = 10$ and $m_2 = 9$). It can be seen that, several angular momentum
components contribute to the reflected wave component $A_l$, and the
asymmetric distribution of the angular momentum components about zero leads to
skew scattering. For the transmitted wave components, the distribution of
the angular components is asymmetric as well, leading to the emergence of
resonant vortices. For the $B_l$ coefficients, their values for the
non-skyrmion case is much smaller than those for the skyrmion case, indicating
that the skyrmion structure can confine the electrons much more effectively
than the non-skyrmion structure.

\begin{figure} [ht!]
\centering
\includegraphics[width=\linewidth]{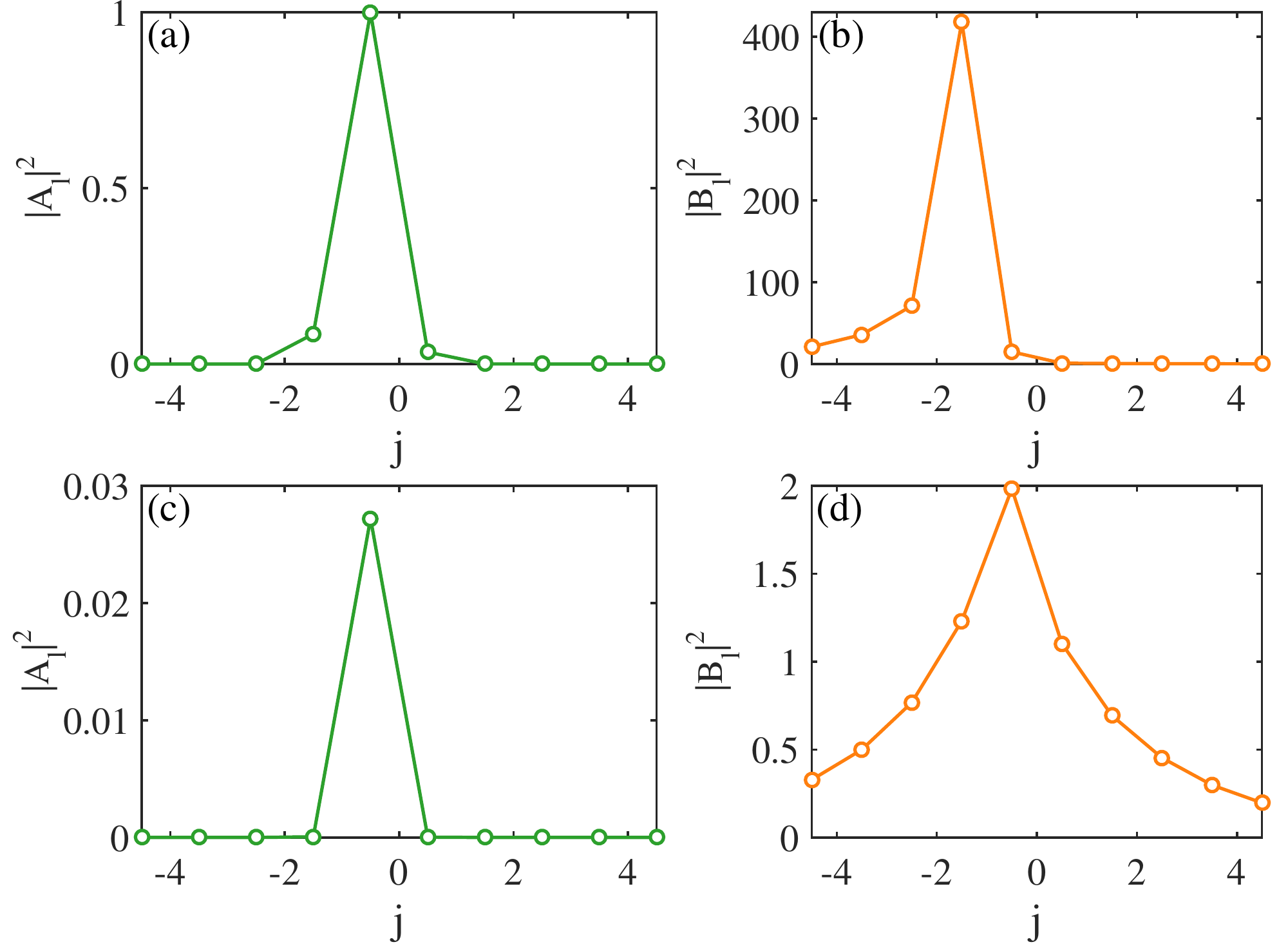}
\caption{ Transmitted and reflected partial wave coefficients as a function
of the total angular momentum for a circular magnetic structure in the long
wavelength regime. The radius of the structure is $R = 0.1$. (a,b) $|A_l|^2$
and $|B_l|^2$ versus $j$ for $m_1 = 10$ and $m_2 = -9$ (skyrmion case),
respectively, where the state is the one shown in
Fig.~\ref{fig:CS-Mass-Circle-StateS}(a). (c,d) $|A_l|^2$ and $|B_l|^2$ versus
$j$ for $m_1 = 10$ and $m_2 = 9$ (non-skyrmion case), respectively, where the
corresponding state is shown in Fig.~\ref{fig:CS-Mass-Circle-StateS}(b).}
\label{fig:AB-SmallR}
\end{figure}

Setting $R = 0.1$ lands the scattering system in the long wavelength regime.
Figures~\ref{fig:AB-SmallR}(a,b) and \ref{fig:AB-SmallR}(c,d) show the
coefficients associated with different angular-momentum components for
the skyrmion ($m_1 = 10$ and $m_2 = -9$) and non-skyrmion ($m_1 = 10$ and
$m_2 = 9$) cases, respectively. In both cases, only a single angular momentum
component contributes to the coefficient $A_l$, i.e., $j = -1/2$, giving rise
to the directionality in the scattering and a slow change in the resonant
cross section with the energy. The value of $A_l$ for the non-skyrmion case is
much smaller than that of the skyrmion case. For the transmitted coefficient
$B_l$, the angular momentum component $j = -3/2$ dominates the skyrmion case
and a number of components including $j =-1/2$ have contributions in the
non-skyrmion case, and the values of $B_l$ are much larger in the skyrmion
than the non-skyrmion case, again implying stronger confinement by resonance
and better directionality of scattering in the skyrmion structure as compared
with those in the non-skyrmion structure.

\section{Discussion} \label{sec:discussion}

We have investigated relativistic quantum scattering of Dirac electrons from
a closed magnetic structure embedded in the top surface of a 3D TI. Outside
of the structure, there is a uniform FMI layer, leading to a finite but
positive mass for the Dirac electron. The mass of the structure itself can be
engineered to be negative or positive, where a skyrmion and a non-skyrmion
structure arises in the former and latter case, respectively. In the short
wavelength regime, the nature of the classical dynamics in the closed
structure should be relevant to the quantum scattering dynamics, according
to conventional wisdom from the study of quantum
chaos~\cite{Stockmann:book,Haake:book}. For a perfectly circular structure,
the classical dynamics are integrable. For a deformed structure such as one
with the stadium shape, there is fully developed chaos in the classical
dynamics. Our main findings are two. Firstly, in the short wavelength regime,
classical chaos hardly has any effect on the scattering dynamics. In fact,
similar behaviors in the scattering characteristics at a {\em quantitative}
level, such as the skew scattering and backscattering cross sections, have
arise for the circular and stadium-shaped structures. The diminishing effects
of classical chaos on relativistic quantum scattering from a magnetic
structure are consistent with previous results on weakened manifestations of
chaos in relativistic quantum systems in
general~\cite{XL:2019,HXHL:2019,YHLG:2011a,YHLG:2012,XHLG:2015}. Secondly,
strong skew scattering can arise when the magnetic structure is a skyrmion,
regardless of the nature of the classical dynamics. In the short wavelength
regime, the pronounced skew scattering is associated with resonant modes
manifested as confined vortices inside of the skyrmion structure, which are
originated from the sign change in the mass when the Dirac electrons travel
from outside to inside of the skyrmion structure. A partial wave analysis
for scattering from a circular skyrmion has revealed that a large number of
angular momentum channels contribute to the resonant modes. We have also
studied the long wavelength regime, where the geometric details of the
magnetic structure are unresolved so naturally the scattering process is
expected to be independent of the nature of the classical dynamics. In this
regime, resonant states can still emerge as confined edge states inside of
the magnetic structure, to which only a single angular momentum channel
contributes, leading to highly directional skew scattering.

In the short wavelength regime, the resonant states manifested as confined
vortices inside of the skyrmion structure can be exploited for electrically
charging the skyrmion structure~\cite{nomura2010electric,hurst2015charged},
enabling the surface electrons on the TI to drive skyrmion motion with a low
current and high thermal efficiency. In the long wavelength regime, the strong
and robust directionality for skew scattering may be exploited for device
application based on the anomalous Hall effect. 

About experimental realization of a skyrmion structure, we note that there is
recent evidence of magnetic skyrmion at the interface of the ferromagnet/TI 
(Cr$_2$Te$_3$/Bi$_2$Te$_3$) heterostructure~\cite{Chenetal:2019}. In 
addition, inhomogeneous Zeeman coupling can be tuned for a ferromagnetic 
strip with strong out-of-plane magnetic anisotropy~\cite{hurst2015charged}.
For experimental control of electron scattering over a skyrmion structure, a 
quantum-dot type of configuration with skyrmion structure in a finite 
scattering region as well as with leads and contacts is necessary. The 
scattering configuration employed in our work is mainly for theoretical 
convenience with the goal to gain insights into the physics of electron 
scattering over the skyrmion structure with classical integrable or chaotic 
dynamics. For this purpose, the geometrical structure of the skyrmion is 
chosen to be either circular for which the scattering cross sections can be 
calculated analytically, or deformed for which the numerical method of 
multiple multipoles can be used to calculate the scattering wave function 
and consequently the resonant states, the cross sections, the current and 
spin distribution. Our results provide useful hints about the 
scattering of spin-1/2 fermion over a skyrmion structure. If the
device size is significantly larger than the electron wavelength, we 
expect the main results to hold.

A number of open issues are
worth studying, such as using spin transfer torque of the electrons to
drive the skyrmion motion, exploitation of skyrmion related switches or
oscillators, and scattering from multiple skyrmions that are themselves
dynamic with possible phase-locking or anti-phase locking behavior.

\begin{acknowledgments}

This work was supported by the Pentagon Vannevar Bush Faculty Fellowship
program sponsored by the Basic Research Office of the Assistant Secretary of
Defense for Research and Engineering and funded by the Office of Naval
Research through Grant No.~N00014-16-1-2828.

\end{acknowledgments}

\appendix

\section*{Appendix: Multiple multipole (MMP) method for scattering of Dirac 
electrons on the top of a TI from a magnetic structure} 

\begin{figure}[htp]
\centering
\includegraphics[width=\linewidth]{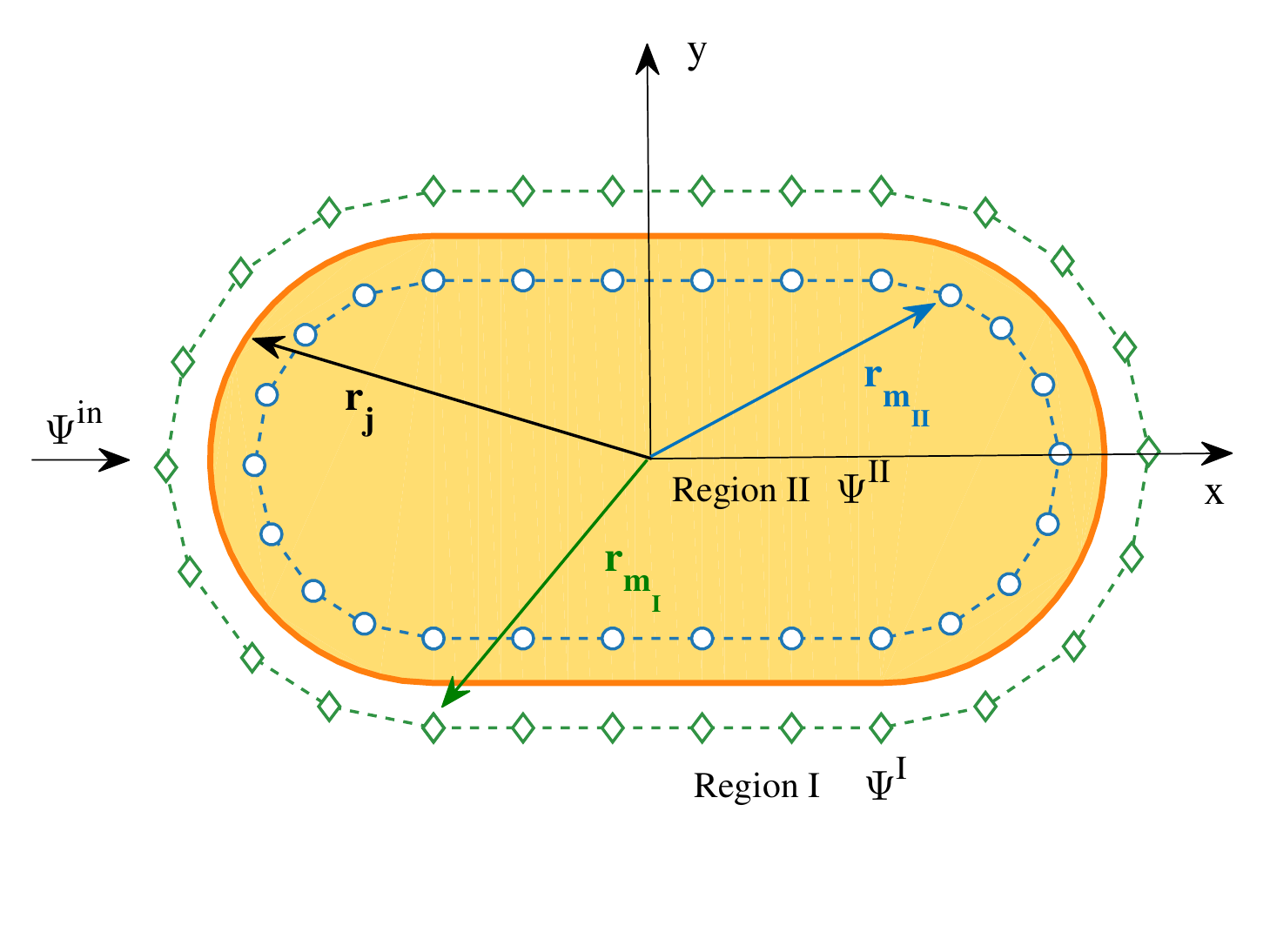}
\caption{ A schematic illustration of the basics of the MMP method. Shown
is placement of poles (fictitious sources) inside and outside of a magnetic
structure of arbitrary shape. The scattering spinor wavefunctions inside
(outside) of the structure are determined by the poles outside (inside) of
the structure.}
\label{fig:Appendix1}
\end{figure}

We denote the area outside and inside of the skyrmion structure as regions
$I$ and $II$, respectively. The wavefunction in region $II$ can be written as
\begin{align}
&\Psi^{II}(\boldsymbol{r})
\equiv
\left(
      \begin{array}{cc}
        \psi^{I}_1\\
        \psi^{II}_2
      \end{array}
\right)\nonumber\\
&= \sum_{m_I} \sum_{l} \frac{C_l^{m_I}}{\sqrt{2}}
\left(
      \begin{array}{cc}
        H^{(1)}_{l-1}(k_{II}d_{m_I})e^{-i\theta_{m_I}} \\
        \tau_{II}H_l^{(1)}(k_{II}d_{m_I})
      \end{array}
\right)e^{il\theta_{m_I}},
\end{align}
where
\begin{eqnarray}
\nonumber
k_{II} & = & \sqrt{E^2 - \Delta^2 n^2_{II}}/\hbar v_F, \\ \nonumber
\tau_{II} & = & -\hbar v_F k_{II}/(E - \Delta n_{II}), \\ \nonumber
d_{m_I} & = & |\boldsymbol{r} - \boldsymbol{r}_{m_I}|, \\ \nonumber
\theta_{m_I} & = & \mbox{Angle}(\boldsymbol{r} - \boldsymbol{r}_{m_I}),
\end{eqnarray}
and $C_l^{m_I}$ are the expansion coefficients. The scattered wavefunction
in region $I$ is
\begin{align}
&\Psi^I(\boldsymbol{r})
\equiv
\left(
      \begin{array}{cc}
        \psi^I_1\\
        \psi^I_2
      \end{array}
\right)   \nonumber\\
&= \sum_{m_{II}} \sum_{l} \frac{C_l^{m_{II}}}{\sqrt{2}}
\left(
      \begin{array}{cc}
        H^{(1)}_{l-1}(k_I d_{m_{II}})e^{-i\theta_{m_{II}}} \\
        \tau_I H_l^{(1)}(k_I d_{m_{II}})
      \end{array}
\right)e^{il\theta_{m_{II}}},
\end{align}
where
\begin{eqnarray}
\nonumber
k_I & = & \sqrt{(E - \Delta^2 n^2_I}/\hbar v_F, \\ \nonumber
\tau_I & = & -\hbar v_F k_I/(E - \Delta n_I), \\ \nonumber
d_{m_{II}} & = & |\boldsymbol{r} - \boldsymbol{r}_{m_{II}}|, \\ \nonumber
\theta_{m_{II}} & = & \mbox{Angle}(\boldsymbol{r} - \boldsymbol{r}_{m_{II}}),
\end{eqnarray}
and $C_l^{m_{II}}$ are the expansion coefficients. The incident plane wave
propagating along the direction defined by an angle $\beta$ with the $x$ axis
in region $I$ is given by
\begin{align}
\Psi^{in}(\boldsymbol{r})
&\equiv
\left(
      \begin{array}{cc}
        \psi_1^{in}\\
        \psi_2^{in}
      \end{array}
\right) \nonumber \\
&= \frac{1}{\sqrt{2}}
\left(
      \begin{array}{cc}
        1 \\
        -i\tau_I e^{i\beta}
      \end{array}
\right)e^{i(k_x r \cos\theta + k_y r \sin\theta)}.
\end{align}
Matching the boundary conditions
\begin{align} \label{Eq:BC}
  (\psi_1^I + \psi_1^{in})|_{r_j \in \Gamma} = \psi_1^{II}|_{r_j \in \Gamma} \\
  (\psi_2^I + \psi_2^{in})|_{r_j \in \Gamma} = \psi_2^{II}|_{r_j \in \Gamma},
\end{align}
we get
\begin{eqnarray} \label{Eq:BC1}
\nonumber
& & \sum_{m_{II}} \sum_{l} C_l^{m_{II}} \frac{1}{\sqrt{2}}\tau_I H_l^{(1)}
  (k_I|\boldsymbol{r}_j - \boldsymbol{r}_{m_{II}}|)
  e^{il\theta_{m_{II}}} \\ \nonumber
& - & \sum_{m_I} \sum_{l} C_l^{m_I} \frac{1}{\sqrt{2}}\tau_{II}H_l^{(1)}
  (k_{II}|\boldsymbol{r}_j - \boldsymbol{r}_{m_I}|)
  e^{il\theta_{m_I}} \\
& = & \frac{i}{\sqrt{2}}\tau_I e^{i\beta}e^{i\boldsymbol{k}_I\boldsymbol{r}}
\end{eqnarray}
and
\begin{eqnarray}\label{Eq:BC2}
& & \sum_{m_{II}} \sum_{l} C_l^{m_{II}} \frac{1}{\sqrt{2}}H_{l-1}^{(1)}
  (k_I|\boldsymbol{r}_j - \boldsymbol{r}_{m_{II}}|)
  e^{i(l-1)\theta_{m_{II}}} \\ \nonumber
& - & \sum_{m_I} \sum_{l} C_l^{m_I} \frac{1}{\sqrt{2}}H_{l-1}^{(1)}
  (k_{II}|\boldsymbol{r}_j - \boldsymbol{r}_{m_I}|)
  e^{i(l-1)\theta_{m_I}} \\
& = & -\frac{1}{\sqrt{2}}\tau_I e^{i\beta}e^{i\boldsymbol{k}_I\boldsymbol{r}},
\end{eqnarray}
which can be cast in a compact form as
\begin{align} \label{Eq:Compact1}
  \sum_{m_{II}} \sum_{l} {}^{j}A_{lm_{II}}^I C_l^{m_{II}} - \sum_{m_I} \sum_{l} {}^{j}A^{II}_{lm_I} C_l^{m_I} = - {}^{j}\psi^{in}_{II}
\end{align}
\begin{align} \label{Eq:Compact2}
  \sum_{m_{II}} \sum_{l} {}^{j}B^I_{lm_{II}} C_l^{m_{II}} - \sum_{m_{I}} \sum_{l} {}^{j}B^{II}_{lm_I} C_l^{m_I} = - {}^{j}\psi^{in}_{I}
\end{align}
where
\begin{align} \label{Eq:matrix-AB}
  {}^{j}A^I_{lm_{II}} &= \frac{1}{\sqrt{2}} \tau_I H^{(1)}_l (k_I |\boldsymbol{r}_j - \boldsymbol{r}_{m_{II}}|)e^{il\theta_{m_{II}}},\\
  {}^{j}A^{II}_{lm_I} &= \frac{1}{\sqrt{2}} \tau_{II} H^{(1)}_l (k_{II} |\boldsymbol{r}_j - \boldsymbol{r}_{m_I}|)e^{il\theta_{m_I}},\\
  {}^{j}B^I_{lm_{II}} &= \frac{1}{\sqrt{2}} H^{(1)}_{l-1} (k_I |\boldsymbol{r}_j - \boldsymbol{r}_{m_{II}}|)e^{i(l-1)\theta_{m_{II}}},\\
  {}^{j}B^{II}_{lm_I} &= \frac{1}{\sqrt{2}} H^{(1)}_{l-1} (k_{II} |\boldsymbol{r}_j - \boldsymbol{r}_{m_I}|)e^{i(l-1)\theta_{m_I}},
\end{align}
and
\begin{align} \label{Eq:matrix-psi}
  {}^{j}\psi^{in}_{2} &= -\frac{i}{\sqrt{2}} \tau_I e^{i\beta} e^{i \boldsymbol{k}_I \boldsymbol{r}_j},\\
  {}^{j}\psi^{in}_{1} &= \frac{1}{\sqrt{2}} e^{i \boldsymbol{k}_I \boldsymbol{r}_j}.
\end{align}
In principle, the set consists of an infinite number of equations with an
infinite number of undetermined expansion coefficients $C^{m_{II}}_l$ and
$C^{m_I}_l$. To solve the system numerically, finite truncation is necessary.
We set the total number of boundary points to be $J$ with $M_I$ and $M_{II}$
poles in regions $I$ and $II$, respectively, and $l \to [-L, L]$ for all the
multipoles. The process leads to the following finite-dimensional matrix
equation:
\begin{align}
  \mathbb{M}_{2J\times N} \cdot C_{N\times 1} = -Y_{2J\times 1},
\end{align}
where $N = (2L + 1)\times (M_I + M_{II}) = N_I + N_{II}$,
\begin{widetext}
\begin{align}
  C_{N \times 1} = \begin{pmatrix}
  C^{1_{II}}_{-L} \\
  \vdots \\
  C^{1_{II}}_{l} \\
  C^{2_{II}}_{l} \\
  \vdots \\
  C^{M_{II}}_{l} \\
  \vdots \\
  C^{M_{II}}_{L} \\
  \\
  \hline
  \\
  C^{1_{I}}_{-L} \\
  \vdots \\
  C^{1_{I}}_{l} \\
  C^{2_{I}}_{l} \\
  \vdots \\
  C^{M_{I}}_{l} \\
  \vdots \\
  C^{M_{I}}_{L} \\
\end{pmatrix}_{N\times 1} \quad ; \quad
  Y_{2J \times 1} = \begin{pmatrix}
  {}^1\psi^{in}_2 \\
  \vdots \\
  {}^j\psi^{in}_2 \\
  \vdots \\
  {}^J\psi^{in}_2 \\
  \\
  \hline
  \\
  {}^1\psi^{in}_1 \\
  \vdots \\
  {}^j\psi^{in}_1 \\
  \vdots \\
  {}^J\psi^{in}_1 \\
\end{pmatrix}_{2J\times 1}
\end{align}
and
\begin{align}
  \mathbb{M}_{2J\times N}=
\left(\begin{array}{c|c}
  \mathbb{A}^{(I)}& -\mathbb{A}^{(II)} \\
  \hline
  \mathbb{B}^{(I)} & -\mathbb{B}^{(II)}
  \end{array}
\right)
\end{align}
with
\begin{equation}\label{Eq:Atau}
\mathbb{A}^{(\tau)} =
\left(
             \begin{array}{cccccccc}
             {}^1A^{(\tau)}_{-L1_{\overline{\tau}}} & \cdots & {}^1A^{(\tau)}_{l1_{\overline{\tau}}}  & {}^1A^{(\tau)}_{l2_{\overline{\tau}}}  & \cdots & {}^1A^{(\tau)}_{lM_{\overline{\tau}}}  & \cdots &
             {}^1A^{(\tau)}_{LM_{\overline{\tau}}} \\
             {}^2A^{(\tau)}_{-L1_{\overline{\tau}}} & \cdots & {}^2A^{(\tau)}_{l1_{\overline{\tau}}}  & {}^2A^{(\tau)}_{l2_{\overline{\tau}}}  & \cdots & {}^2A^{(\tau)}_{lM_{\overline{\tau}}}  & \cdots &
             {}^2A^{(\tau)}_{LM_{\overline{\tau}}} \\
             \vdots & \cdots & \vdots  & \vdots & \cdots & \vdots  & \cdots & \vdots \\
             {}^jA^{(\tau)}_{-L1_{\overline{\tau}}} & \cdots & {}^jA^{(\tau)}_{l1_{\overline{\tau}}}  & {}^jA^{(\tau)}_{l2_{\overline{\tau}}}  & \cdots & {}^jA^{(\tau)}_{lM_{\overline{\tau}}}  & \cdots &
             {}^jA^{(\tau)}_{LM_{\overline{\tau}}} \\
             \vdots & \cdots & \vdots  & \vdots & \cdots & \vdots  & \cdots & \vdots \\
             {}^JA^{(\tau)}_{-L1_{\overline{\tau}}} & \cdots & {}^JA^{(\tau)}_{l1_{\overline{\tau}}}  & {}^JA^{(\tau)}_{l2_{\overline{\tau}}}  & \cdots & {}^JA^{(\tau)}_{lM_{\overline{\tau}}}  & \cdots &
             {}^JA^{(\tau)}_{LM_{\overline{\tau}}}
             \end{array}
\right).
\end{equation}

\begin{equation}\label{Eq:Btau}
\mathbb{B}^{(\tau)} =
\left(
             \begin{array}{cccccccc}
             {}^1B^{(\tau)}_{-L1_{\overline{\tau}}} & \cdots & {}^1B^{(\tau)}_{l1_{\overline{\tau}}}  & {}^1B^{(\tau)}_{l2_{\overline{\tau}}}  & \cdots & {}^1B^{(\tau)}_{lM_{\overline{\tau}}}  & \cdots &
             {}^1B^{(\tau)}_{LM_{\overline{\tau}}} \\
             {}^2B^{(\tau)}_{-L1_{\overline{\tau}}} & \cdots & {}^2B^{(\tau)}_{l1_{\overline{\tau}}}  & {}^2B^{(\tau)}_{l2_{\overline{\tau}}}  & \cdots & {}^2B^{(\tau)}_{lM_{\overline{\tau}}}  & \cdots &
             {}^2B^{(\tau)}_{LM_{\overline{\tau}}} \\
             \vdots & \cdots & \vdots  & \vdots & \cdots & \vdots  & \cdots & \vdots \\
             {}^jB^{(\tau)}_{-L1_{\overline{\tau}}} & \cdots & {}^jB^{(\tau)}_{l1_{\overline{\tau}}}  & {}^jB^{(\tau)}_{l2_{\overline{\tau}}}  & \cdots & {}^jB^{(\tau)}_{lM_{\overline{\tau}}}  & \cdots &
             {}^jB^{(\tau)}_{LM_{\overline{\tau}}} \\
             \vdots & \cdots & \vdots  & \vdots & \cdots & \vdots  & \cdots & \vdots \\
             {}^JB^{(\tau)}_{-L1_{\overline{\tau}}} & \cdots & {}^JB^{(\tau)}_{l1_{\overline{\tau}}}  & {}^JB^{(\tau)}_{l2_{\overline{\tau}}}  & \cdots & {}^JB^{(\tau)}_{lM_{\overline{\tau}}}  & \cdots &
             {}^JB^{(\tau)}_{LM_{\overline{\tau}}}
             \end{array}
\right).
\end{equation}
\end{widetext}
As the expansions are generally nonorthogonal, more equations are required
than the number of unknowns to enable reduction of an overdetermined matrix
system with $2J \gg N$, which can be solved by the pseudoinverse algorithm:
$\boldsymbol{C} = -\mbox{pinv}(\mathbb{M}*\boldsymbol{Y})$.


%
\end{document}